\documentclass[10pt,reqno]{article}

\renewcommand{\refname}{\textbf{References}}

\usepackage{graphicx}
\usepackage{amssymb,amsmath,amsthm,amscd}
\usepackage{enumerate}
\usepackage{color}
\usepackage[dvipsnames]{xcolor}
\usepackage{verbatim}
\usepackage{epsfig}
\usepackage{latexsym}
\usepackage{texdraw}
\usepackage{amsmath, amssymb, latexsym}
\usepackage{amsfonts}
\usepackage[mathscr]{eucal}
\usepackage{graphics}
\DeclareGraphicsExtensions{.pdf,.png,.jpg}
\usepackage{mathrsfs}
\usepackage[all,knot]{xy}
\usepackage{mathabx}
\usepackage{MnSymbol}
\usepackage{tikz}
\usepackage{rotating}
\usepackage{longtable}
\usepackage{hyperref}
\usepackage[numbered]{bookmark}

\usepackage{tabu}
\usepackage{booktabs}

\usetikzlibrary{shapes}
\xyoption{arc}

\theoremstyle{plain}

\theoremstyle{definition}
\newtheorem{exa}{Example}
\theoremstyle{remark}
\newtheorem{case}{Case}

\newcommand{\G}{{\mathcal G}}
\newcommand{\C}{{\chi}}
\newcommand{\OC}{{\chi_{orb}}}
\newcommand{\underX}{{\lvert X \rvert}}
\newcommand{\underIX}{{\lvert IX \rvert}}
\newcommand{\D}{{\bigsqcup}}

\title{A Short Note on Improved Logic Circuits in a Hexagonal Minesweeper}
\date{\today}
\author{Seunghoon Lee\\Department of Mathematics, Seoul National University\footnote{Currently on a leave of absence for the mandatory military service.}\\galaxybp@snu.ac.kr}

\begin{document}

\maketitle

%%%%%%%%%%%%%%%%%%%%%%%
%%%%% 0. Abstract %%%%%
%%%%%%%%%%%%%%%%%%%%%%%

\begin{abstract}
\noindent This paper aims to present an advanced version of PP-hardness proof of Minesweeper by Bondt \cite{bondt12}. The advancement includes improved Minesweeper configurations for `logic circuits' in a hexagonal Minesweeper. To do so, I demonstrate logical uncertainty in Minesweeper, which ironically allows a possibility to make some Boolean operators.\\
\indent The fact that existing hexagonal logic circuits did not clearly distinguish the {\textit{true}} and {\textit{false}} signal needs an improved form of a hexagonal wire. I introduce new forms of logic circuits such as NOT, AND, OR gates, a curve and a splitter of wires. Moreover, these new logic circuits complement Bondt's \cite{bondt12} proof for PP-hardness of Minesweeper by giving a new figure.\\
   
\end{abstract}

%%%%%%%%%%%%%%%%%%%%%%%%%%%
%%%%% 1. Introduction %%%%%
%%%%%%%%%%%%%%%%%%%%%%%%%%%

\section{Introduction}

Every computer user in the world must, at least once, have played this game: Minesweeper. If anyone ever witnessed the yellow circle smiling with its sunglasses on, she or he may find this research interesting. A normal Minesweeper is played on a square grid, each compartment is enclosed by eight neighborhoods except on the border of the grid. When we click any compartment on the grid, a number between 0 and 8 appears showing the number of mines around the point one clicked. Of course, if the clicked point was exactly on a mine, then the game is over. The goal of Minesweeper is to find and check all compartments, which contain mines. There are many strategies to win the game, but existing strategies do not provide a perfect logic to beat the game. The following example shows this uncertainty.

\begin{exa}[Uncertainty on $9\times 9$ board]
During Minesweeper gameplay in novice mode which consists of a $9\times 9$ board, let us be given the situation as figure~\hyperref[fig:example]{\ref{fig:example}(a)}.

%%%%%%%%%%%%%%%%%%%%%%
% Figure 1 : Example %
%%%%%%%%%%%%%%%%%%%%%%

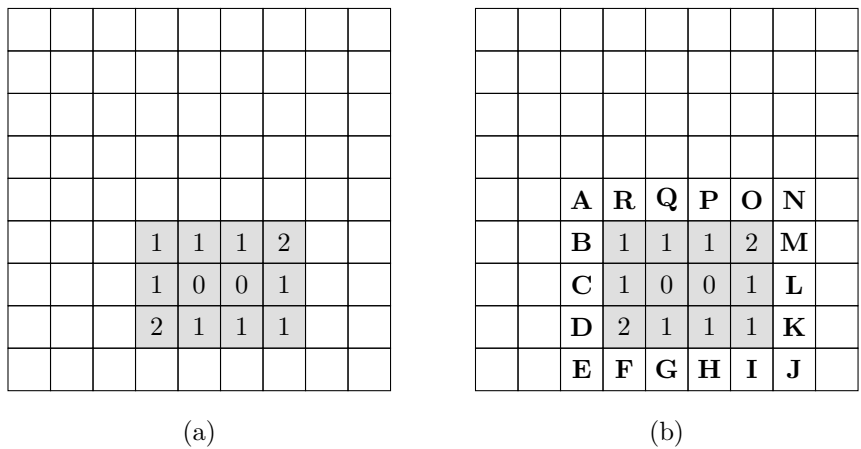
\begin{figure}[ht]
\centering
\begin{tikzpicture} [rect/.style= {shape=regular polygon,regular polygon sides=4,minimum size=0.8cm, draw,inner sep=0,anchor=south}]

% fill in some colors

\foreach \k in {3,4,5,6,14,15,16,17}{
  \foreach \j in {1,2,3}
  {\node [shape=regular polygon,regular polygon sides=4,minimum size=0.8cm,fill=lightgray!50!white,inner sep=0,anchor=south] at ({(\k)*0.565},{(\j)*0.565}) {};} }

% generating nodes

\foreach \j in {0,...,8}{ 
  \foreach \i in {0,1,2,3,4,5,6,7,8,11,12,13,14,15,16,17,18,19}{
  \node[rect] (h\i;\j) at ({(\i)*0.565},{\j*0.565}) {};}  }      

% putting in contents

\foreach \k in {4,5,6,15,16,17} {\node at (h\k;1) {1};}
\foreach \k in {3,14} {\node at (h\k;1) {2};}

\foreach \k in {4,5,15,16} {\node at (h\k;2) {0};}
\foreach \k in {3,6,14,17} {\node at (h\k;2) {1};}

\foreach \k in {3,4,5,14,15,16} {\node at (h\k;3) {1};}
\foreach \k in {6,17} {\node at (h\k;3) {2};}

{\node at ({4*0.565},{-1*0.565}) {(a)};}
{\node at ({15*0.565},{-1*0.565}) {(b)};}

{\node at (h13;0) {\textbf{E}};}
{\node at (h14;0) {\textbf{F}};}
{\node at (h15;0) {\textbf{G}};}
{\node at (h16;0) {\textbf{H}};}
{\node at (h17;0) {\textbf{I}};}
{\node at (h18;0) {\textbf{J}};}
{\node at (h13;4) {\textbf{A}};}
{\node at (h14;4) {\textbf{R}};}
{\node at (h15;4) {\textbf{Q}};}
{\node at (h16;4) {\textbf{P}};}
{\node at (h17;4) {\textbf{O}};}
{\node at (h18;4) {\textbf{N}};}
{\node at (h13;3) {\textbf{B}};}
{\node at (h13;2) {\textbf{C}};}
{\node at (h13;1) {\textbf{D}};}
{\node at (h18;3) {\textbf{M}};}
{\node at (h18;2) {\textbf{L}};}
{\node at (h18;1) {\textbf{K}};}

\end{tikzpicture}
\caption{An example that shows uncertainty of Minesweeper}\label{fig:example}
\end{figure}

At first glance, it is difficult to determine the locations of mines. Once we allocate names from {\textbf{A}} to {\textbf{R}} as shown in \hyperref[fig:example]{(b)}. Then, there are 30 different possible locations for the mines. For example, if we set {\textbf{B}} as a mine, then {\textbf{C,D,A,R,Q}} should not be mines. Therefore, {\textbf{P}} should be a mine; since two of {\textbf{E,F,G}}, one of {\textbf{F,G,H}}, and one of {\textbf{G,H,I}} should be a mine, we conclude that {\textbf{E}} is a mine. Next choice is either {\textbf{F}} or {\textbf{G}}, if {\textbf{F}} is a mine, then automatically so are {\textbf{I}} and {\textbf{M}}. All possible arrangements of mines are given in the table~\ref{tab:table1}\footnote{$\ovoid$ means identified as a mine, and $\bigtimes$ means there is no mine.}:

\begin{longtabu} to \textwidth { |X|X|X|X|X|X|X|X|X|X|X|X|X|X|X|X|X|X|X| }
\caption{30 different possibilities}\\
	\hline
     & \bf{B} & \bf{C} & \bf{D} & \bf{E} & \bf{F} & \bf{G} & \bf{H} & \bf{I} & \bf{J} & \bf{K} & \bf{L} & \bf{M} & \bf{N} & \bf{O} & \bf{P} & \bf{Q} & \bf{R} & \bf{A} \\
    \hline\hline
    \endhead
    1 & $\ovoid$ & $\bigtimes$ & $\bigtimes$ & $\ovoid$ & $\ovoid$ & $\bigtimes$ & $\bigtimes$ & $\ovoid$ & $\bigtimes$ & $\bigtimes$ & $\bigtimes$ & $\ovoid$ & $\bigtimes$ & $\bigtimes$ & $\ovoid$ & $\bigtimes$ & $\bigtimes$ & $\bigtimes$ \\\hline
    2 & $\ovoid$ & $\bigtimes$ & $\bigtimes$ & $\ovoid$ & $\bigtimes$ & $\ovoid$ & $\bigtimes$ & $\bigtimes$ & $\ovoid$ & $\bigtimes$ & $\bigtimes$ & $\ovoid$ & $\bigtimes$ & $\bigtimes$ & $\ovoid$ & $\bigtimes$ & $\bigtimes$ & $\bigtimes$ \\\hline
    3 & $\ovoid$ & $\bigtimes$ & $\bigtimes$ & $\ovoid$ & $\bigtimes$ & $\ovoid$ & $\bigtimes$ & $\bigtimes$ & $\bigtimes$ & $\ovoid$ & $\bigtimes$ & $\bigtimes$ & $\ovoid$ & $\bigtimes$ & $\ovoid$ & $\bigtimes$ & $\bigtimes$ & $\bigtimes$ \\\hline
    4 & $\ovoid$ & $\bigtimes$ & $\bigtimes$ & $\ovoid$ & $\bigtimes$ & $\ovoid$ & $\bigtimes$ & $\bigtimes$ & $\bigtimes$ & $\bigtimes$ & $\ovoid$ & $\bigtimes$ & $\bigtimes$ & $\bigtimes$ & $\ovoid$ & $\bigtimes$ & $\bigtimes$ & $\bigtimes$ \\\hline
    5 & $\bigtimes$ & $\ovoid$ & $\bigtimes$ & $\ovoid$ & $\bigtimes$ & $\bigtimes$ & $\ovoid$ & $\bigtimes$ & $\bigtimes$ & $\bigtimes$ & $\bigtimes$ & $\ovoid$ & $\bigtimes$ & $\bigtimes$ & $\ovoid$ & $\bigtimes$ & $\bigtimes$ & $\bigtimes$ \\\hline
    6 & $\bigtimes$ & $\ovoid$ & $\bigtimes$ & $\bigtimes$ & $\ovoid$ & $\bigtimes$ & $\bigtimes$ & $\ovoid$ & $\bigtimes$ & $\bigtimes$ & $\bigtimes$ & $\ovoid$ & $\bigtimes$ & $\bigtimes$ & $\ovoid$ & $\bigtimes$ & $\bigtimes$ & $\bigtimes$ \\\hline
    7 & $\bigtimes$ & $\ovoid$ & $\bigtimes$ & $\bigtimes$ & $\bigtimes$ & $\ovoid$ & $\bigtimes$ & $\bigtimes$ & $\ovoid$ & $\bigtimes$ & $\bigtimes$ & $\ovoid$ & $\bigtimes$ & $\bigtimes$ & $\ovoid$ & $\bigtimes$ & $\bigtimes$ & $\bigtimes$ \\\hline
    8 & $\bigtimes$ & $\ovoid$ & $\bigtimes$ & $\bigtimes$ & $\bigtimes$ & $\ovoid$ & $\bigtimes$ & $\bigtimes$ & $\bigtimes$ & $\ovoid$ & $\bigtimes$ & $\bigtimes$ & $\ovoid$ & $\bigtimes$ & $\ovoid$ & $\bigtimes$ & $\bigtimes$ & $\bigtimes$ \\\hline
    9 & $\bigtimes$ & $\ovoid$ & $\bigtimes$ & $\bigtimes$ & $\bigtimes$ & $\ovoid$ & $\bigtimes$ & $\bigtimes$ & $\bigtimes$ & $\bigtimes$ & $\ovoid$ & $\bigtimes$ & $\bigtimes$ & $\bigtimes$ & $\ovoid$ & $\bigtimes$ & $\bigtimes$ & $\bigtimes$ \\\hline
    10 & $\bigtimes$ & $\bigtimes$ & $\ovoid$ & $\ovoid$ & $\bigtimes$ & $\bigtimes$ & $\ovoid$ & $\bigtimes$ & $\bigtimes$ & $\bigtimes$ & $\bigtimes$ & $\ovoid$ & $\ovoid$ & $\bigtimes$ & $\bigtimes$ & $\ovoid$ & $\bigtimes$ & $\bigtimes$ \\\hline
    11 & $\bigtimes$ & $\bigtimes$ & $\ovoid$ & $\ovoid$ & $\bigtimes$ & $\bigtimes$ & $\ovoid$ & $\bigtimes$ & $\bigtimes$ & $\bigtimes$ & $\bigtimes$ & $\ovoid$ & $\bigtimes$ & $\ovoid$ & $\bigtimes$ & $\bigtimes$ & $\ovoid$ & $\bigtimes$ \\\hline
    12 & $\bigtimes$ & $\bigtimes$ & $\ovoid$ & $\ovoid$ & $\bigtimes$ & $\bigtimes$ & $\ovoid$ & $\bigtimes$ & $\bigtimes$ & $\bigtimes$ & $\bigtimes$ & $\ovoid$ & $\bigtimes$ & $\ovoid$ & $\bigtimes$ & $\bigtimes$ & $\bigtimes$ & $\ovoid$ \\\hline
    13 & $\bigtimes$ & $\bigtimes$ & $\ovoid$ & $\ovoid$ & $\bigtimes$ & $\bigtimes$ & $\ovoid$ & $\bigtimes$ & $\bigtimes$ & $\bigtimes$ & $\bigtimes$ & $\ovoid$ & $\bigtimes$ & $\bigtimes$ & $\ovoid$ & $\bigtimes$ & $\bigtimes$ & $\ovoid$ \\\hline
    14 & $\bigtimes$ & $\bigtimes$ & $\ovoid$ & $\bigtimes$ & $\ovoid$ & $\bigtimes$ & $\bigtimes$ & $\ovoid$ & $\bigtimes$ & $\bigtimes$ & $\bigtimes$ & $\ovoid$ & $\ovoid$ & $\bigtimes$ & $\bigtimes$ & $\ovoid$ & $\bigtimes$ & $\bigtimes$ \\\hline
    15 & $\bigtimes$ & $\bigtimes$ & $\ovoid$ & $\bigtimes$ & $\ovoid$ & $\bigtimes$ & $\bigtimes$ & $\ovoid$ & $\bigtimes$ & $\bigtimes$ & $\bigtimes$ & $\ovoid$ & $\bigtimes$ & $\ovoid$ & $\bigtimes$ & $\bigtimes$ & $\ovoid$ & $\bigtimes$ \\\hline
    16 & $\bigtimes$ & $\bigtimes$ & $\ovoid$ & $\bigtimes$ & $\ovoid$ & $\bigtimes$ & $\bigtimes$ & $\ovoid$ & $\bigtimes$ & $\bigtimes$ & $\bigtimes$ & $\ovoid$ & $\bigtimes$ & $\ovoid$ & $\bigtimes$ & $\bigtimes$ & $\bigtimes$ & $\ovoid$ \\\hline
    17 & $\bigtimes$ & $\bigtimes$ & $\ovoid$ & $\bigtimes$ & $\ovoid$ & $\bigtimes$ & $\bigtimes$ & $\ovoid$ & $\bigtimes$ & $\bigtimes$ & $\bigtimes$ & $\ovoid$ & $\bigtimes$ & $\bigtimes$ & $\ovoid$ & $\bigtimes$ & $\bigtimes$ & $\ovoid$ \\\hline
    18 & $\bigtimes$ & $\bigtimes$ & $\ovoid$ & $\bigtimes$ & $\bigtimes$ & $\ovoid$ & $\bigtimes$ & $\bigtimes$ & $\ovoid$ & $\bigtimes$ & $\bigtimes$ & $\ovoid$ & $\ovoid$ & $\bigtimes$ & $\bigtimes$ & $\ovoid$ & $\bigtimes$ & $\bigtimes$ \\\hline
    19 & $\bigtimes$ & $\bigtimes$ & $\ovoid$ & $\bigtimes$ & $\bigtimes$ & $\ovoid$ & $\bigtimes$ & $\bigtimes$ & $\ovoid$ & $\bigtimes$ & $\bigtimes$ & $\ovoid$ & $\bigtimes$ & $\ovoid$ & $\bigtimes$ & $\bigtimes$ & $\ovoid$ & $\bigtimes$ \\\hline
    20 & $\bigtimes$ & $\bigtimes$ & $\ovoid$ & $\bigtimes$ & $\bigtimes$ & $\ovoid$ & $\bigtimes$ & $\bigtimes$ & $\ovoid$ & $\bigtimes$ & $\bigtimes$ & $\ovoid$ & $\bigtimes$ & $\ovoid$ & $\bigtimes$ & $\bigtimes$ & $\bigtimes$ & $\ovoid$ \\\hline
    21 & $\bigtimes$ & $\bigtimes$ & $\ovoid$ & $\bigtimes$ & $\bigtimes$ & $\ovoid$ & $\bigtimes$ & $\bigtimes$ & $\ovoid$ & $\bigtimes$ & $\bigtimes$ & $\ovoid$ & $\bigtimes$ & $\bigtimes$ & $\ovoid$ & $\bigtimes$ & $\bigtimes$ & $\ovoid$ \\\hline
    22 & $\bigtimes$ & $\bigtimes$ & $\ovoid$ & $\bigtimes$ & $\bigtimes$ & $\ovoid$ & $\bigtimes$ & $\bigtimes$ & $\bigtimes$ & $\ovoid$ & $\bigtimes$ & $\bigtimes$ & $\ovoid$ & $\ovoid$ & $\bigtimes$ & $\bigtimes$ & $\ovoid$ & $\bigtimes$ \\\hline
    23 & $\bigtimes$ & $\bigtimes$ & $\ovoid$ & $\bigtimes$ & $\bigtimes$ & $\ovoid$ & $\bigtimes$ & $\bigtimes$ & $\bigtimes$ & $\ovoid$ & $\bigtimes$ & $\bigtimes$ & $\ovoid$ & $\ovoid$ & $\bigtimes$ & $\bigtimes$ & $\bigtimes$ & $\ovoid$ \\\hline
    24 & $\bigtimes$ & $\bigtimes$ & $\ovoid$ & $\bigtimes$ & $\bigtimes$ & $\ovoid$ & $\bigtimes$ & $\bigtimes$ & $\bigtimes$ & $\ovoid$ & $\bigtimes$ & $\bigtimes$ & $\ovoid$ & $\bigtimes$ & $\ovoid$ & $\bigtimes$ & $\bigtimes$ & $\ovoid$ \\\hline
    25 & $\bigtimes$ & $\bigtimes$ & $\ovoid$ & $\bigtimes$ & $\bigtimes$ & $\ovoid$ & $\bigtimes$ & $\bigtimes$ & $\bigtimes$ & $\bigtimes$ & $\ovoid$ & $\bigtimes$ & $\ovoid$ & $\ovoid$ & $\bigtimes$ & $\bigtimes$ & $\ovoid$ & $\bigtimes$ \\\hline
    26 & $\bigtimes$ & $\bigtimes$ & $\ovoid$ & $\bigtimes$ & $\bigtimes$ & $\ovoid$ & $\bigtimes$ & $\bigtimes$ & $\bigtimes$ & $\bigtimes$ & $\ovoid$ & $\bigtimes$ & $\ovoid$ & $\ovoid$ & $\bigtimes$ & $\bigtimes$ & $\bigtimes$ & $\ovoid$ \\\hline
    27 & $\bigtimes$ & $\bigtimes$ & $\ovoid$ & $\bigtimes$ & $\bigtimes$ & $\ovoid$ & $\bigtimes$ & $\bigtimes$ & $\bigtimes$ & $\bigtimes$ & $\ovoid$ & $\bigtimes$ & $\ovoid$ & $\bigtimes$ & $\ovoid$ & $\bigtimes$ & $\bigtimes$ & $\ovoid$ \\\hline
    28 & $\bigtimes$ & $\bigtimes$ & $\ovoid$ & $\bigtimes$ & $\bigtimes$ & $\ovoid$ & $\bigtimes$ & $\bigtimes$ & $\bigtimes$ & $\bigtimes$ & $\ovoid$ & $\bigtimes$ & $\bigtimes$ & $\ovoid$ & $\bigtimes$ & $\bigtimes$ & $\ovoid$ & $\bigtimes$ \\\hline
    29 & $\bigtimes$ & $\bigtimes$ & $\ovoid$ & $\bigtimes$ & $\bigtimes$ & $\ovoid$ & $\bigtimes$ & $\bigtimes$ & $\bigtimes$ & $\bigtimes$ & $\ovoid$ & $\bigtimes$ & $\bigtimes$ & $\ovoid$ & $\bigtimes$ & $\bigtimes$ & $\bigtimes$ & $\ovoid$ \\\hline
    30 & $\bigtimes$ & $\bigtimes$ & $\ovoid$ & $\bigtimes$ & $\bigtimes$ & $\ovoid$ & $\bigtimes$ & $\bigtimes$ & $\bigtimes$ & $\bigtimes$ & $\ovoid$ & $\bigtimes$ & $\bigtimes$ & $\bigtimes$ & $\ovoid$ & $\bigtimes$ & $\bigtimes$ & $\ovoid$ \\\hline
\end{longtabu} \label{tab:table1}

Furthermore, when we check each column({\bf{A}} to {\bf{R}}), we can easily find out that no column contains neither only $\ovoid$ nor only $\bigtimes$; it follows that we cannot go even one step further logically.
\qed
\end{exa}

Using this uncertainty, Kaye \cite{kaye00} made some Minesweeper configurations for `logic circuits', and he proved that Minesweeper is NP-complete. Kaye \cite{kaye00} designated wires carrying either {\textit{true}} or {\textit{false}}; and using these wires he made several logic circuits such as NOT, AND, OR, XOR gates, etc.

In this paper, I apply this concept to a hexagonal grid. In section~\ref{sec3}, I define a hexagonal wire and some logic circuits\footnote{I introduced this concept in a science essay that won President Science scholarship, Republic of Korea in 2004.}. Consequently, this improves Bondt's \cite{bondt12} computational components on a hexagonal Minesweeper.

\newpage

%%%%%%%%%%%%%%%%%%%%%%%%%%%%%%%%%%%%%%%%%%%%%%%%%%%%%%%%%%%%%%%
%%%%% 2. Computational components in a normal Minesweeper %%%%%
%%%%%%%%%%%%%%%%%%%%%%%%%%%%%%%%%%%%%%%%%%%%%%%%%%%%%%%%%%%%%%%

\section{Computational components in a normal Mine-\\sweeper}

First, I will review Kaye's \cite{kaye00} work on Minesweeper configurations. In Figure~\ref{fig:classicwire}, Kaye \cite{kaye00} demonstrated a `wire' that conducts $x$. We can easily figure out that either all $x$'s are 1 and all $x'$'s are 0 or vice versa. It is natural to say the Boolean values \textit{negate} or \textit{confirm} with $x$'s are 0 or 1 respectively.

%%%%%%%%%%%%%%%%%%%%%%%%%%%%%
% Figure 2 : classical wire %
%%%%%%%%%%%%%%%%%%%%%%%%%%%%%

\begin{figure}[ht]
\centering
$\boldsymbol{x} \xrightarrow{\hspace*{2cm}}$
\vskip 0.2em
\begin{tikzpicture} [rect/.style= {shape=regular polygon,regular polygon sides=4,minimum size=0.8cm, draw,inner sep=0,anchor=south}]

% fill in some colors

\foreach \k in {1,3,4,6,7,9,10,12,13,15,16} {\node [shape=regular polygon,regular polygon sides=4,minimum size=0.8cm,fill=lightgray!50!white,inner sep=0,anchor=south] at ({(\k)*0.565},{2*0.565}) {};} 

% generating nodes

\foreach \j in {0,...,4}{ 
  \foreach \i in {0,...,17}{
  \node[rect] (h\i;\j) at ({(\i)*0.565},{\j*0.565}) {};}  }      

% putting in contents

\foreach \k in {1,...,16} {
  \foreach \j in {0,4}
  {\node at (h\k;\j) {0};} }
\foreach \k in {1,...,16} {
  \foreach \j in {1,3}
  {\node at (h\k;\j) {1};} }
\foreach \k in {2,5,8,11,14} {\node at (h\k;2) {1};}
\foreach \k in {1,4,7,10,13,16} {\node at (h\k;2) {$\boldsymbol{x}$};}
\foreach \k in {3,6,9,12,15} {\node at (h\k;2) {$\boldsymbol{x'}$};}

\foreach \k in {0,...,4}  {\node at (h0;\k) {$\cdots$};}
\foreach \k in {0,...,4}  {\node at (h17;\k) {$\cdots$};}

\end{tikzpicture}
\caption{A Wire on a normal Minesweeper}\label{fig:classicwire}
\end{figure}

\begin{figure}[ht]
\centering
\begin{tikzpicture} [rect/.style= {shape=regular polygon,regular polygon sides=4,minimum size=0.8cm, draw,inner sep=0,anchor=south}]

% fill in some colors

\foreach \k in {4,14}{
  \foreach \j in {3,4}
  {\node [shape=regular polygon,regular polygon sides=4,minimum size=0.8cm,fill=lightgray!50!white,inner sep=0,anchor=south] at ({(\k)*0.565},{\j*0.565}) {};} }
\foreach \k in {1,2,11,12}  {\node [shape=regular polygon,regular polygon sides=4,minimum size=0.8cm,fill=lightgray!50!white,inner sep=0,anchor=south] at ({(\k)*0.565},{6*0.565}) {};}
\foreach \k in {8,9}  {\node [shape=regular polygon,regular polygon sides=4,minimum size=0.8cm,fill=lightgray!50!white,inner sep=0,anchor=south] at ({14*0.565},{\k*0.565}) {};}

% generating nodes

\foreach \j in {1,2,3}{ 
  \foreach \i in {3,4,5,13,14,15}{
  \node[rect] (h\i;\j) at ({(\i)*0.565},{\j*0.565}) {};}  }
\foreach \i in {3,4,5,6,13,14,15,16}  {\node[rect] (h\i;4) at ({(\i)*0.565},{4*0.565}) {};}
\foreach \j in {5,6}{ 
  \foreach \i in {0,1,2,3,4,5,6,10,11,12,13,14,15,16}{
  \node[rect] (h\i;\j) at ({(\i)*0.565},{\j*0.565}) {};}  }
\foreach \i in {0,1,2,3,4,5,10,11,12,13,14,15,16}  {\node[rect] (h\i;7) at ({(\i)*0.565},{7*0.565}) {};}
\foreach \i in {2,3,4,13,14,15,16}  {\node[rect] (h\i;8) at ({(\i)*0.565},{8*0.565}) {};}
\foreach \j in {9,10,11}{ 
  \foreach \i in {13,14,15}{
  \node[rect] (h\i;\j) at ({(\i)*0.565},{\j*0.565}) {};}  }

% putting in contents

\foreach \k in {3,4,5,13,14,15} {\node at (h\k;1) {$\vdots$};}
\foreach \k in {3,4,5,13,14,15} {\node at (h\k;2) {1};}
\foreach \k in {3,5,13,15} {\node at (h\k;3) {1};}
\foreach \k in {4,14} {\node at (h\k;3) {$\boldsymbol{x}$};}

\foreach \k in {4,14} {\node at (h\k;4) {$\boldsymbol{x'}$};}
\foreach \k in {6,16} {\node at (h\k;4) {1};}
\foreach \k in {3,13} {\node at (h\k;4) {2};}
\foreach \k in {5,15} {\node at (h\k;4) {3};}

\foreach \k in {0,10}{ \foreach \j in {5,6,7} {\node at (h\k;\j) {$\cdots$};}}

\foreach \k in {1,6,11,16} {\node at (h\k;5) {1};}
\foreach \k in {2,12} {\node at (h\k;5) {2};}
\foreach \k in {3,13} {\node at (h\k;5) {4};}
\foreach \k in {4,5,14,15}  {\node [circle,draw,fill=gray,inner sep=3.5] at (h\k;5) {};}

\foreach \k in {3,4,13,14}  {\node [circle,draw,fill=gray,inner sep=3.5] at (h\k;6) {};}
{\node at (h6;6) {1};}
{\node at (h16;6) {2};}
{\node at (h5;6) {3};}
{\node at (h15;6) {5};}
\foreach \k in {2,12} {\node at (h\k;6) {$\boldsymbol{x}$};}
\foreach \k in {1,11} {\node at (h\k;6) {$\boldsymbol{x'}$};}

\foreach \k in {1,5,11,16} {\node at (h\k;7) {1};}
\foreach \k in {2,4} {\node at (h\k;7) {3};}
{\node at (h12;7) {2};}
{\node at (h13;7) {4};}
\foreach \k in {3,14,15}  {\node [circle,draw,fill=gray,inner sep=3.5] at (h\k;7) {};}

\foreach \k in {2,3,4,16} {\node at (h\k;8) {1};}
{\node at (h13;8) {2};}
{\node at (h15;8) {3};}
{\node at (h14;8) {$\boldsymbol{x'}$};}
{\node at (h14;9) {$\boldsymbol{x}$};}
\foreach \k in {13,15} {\node at (h\k;9) {1};}
\foreach \k in {13,14,15} {\node at (h\k;10) {1};}
\foreach \k in {13,14,15} {\node at (h\k;11) {$\vdots$};}

% arrows

{\node at ({0*0.565},{4.5*0.565}) {$\boldsymbol{x}$};}
\draw [->] ({0.3*0.565},{4.5*0.565}) to [bend left] ({2.1*0.565},{2.5*0.565});

{\node at ({10*0.565},{4.5*0.565}) {$\boldsymbol{x}$};}
\draw [->] ({10.3*0.565},{4.5*0.565}) to [bend left] ({12.1*0.565},{2.5*0.565});
{\node at ({10*0.565},{8.5*0.565}) {$\boldsymbol{x}$};}
\draw [->] ({10.3*0.565},{8.5*0.565}) to [bend right] ({12.1*0.565},{10.5*0.565});

\end{tikzpicture}
\caption{Curve and splitter of a wire (adopted from Bondt \cite{bondt12})}\label{fig:classicbs}
\end{figure}
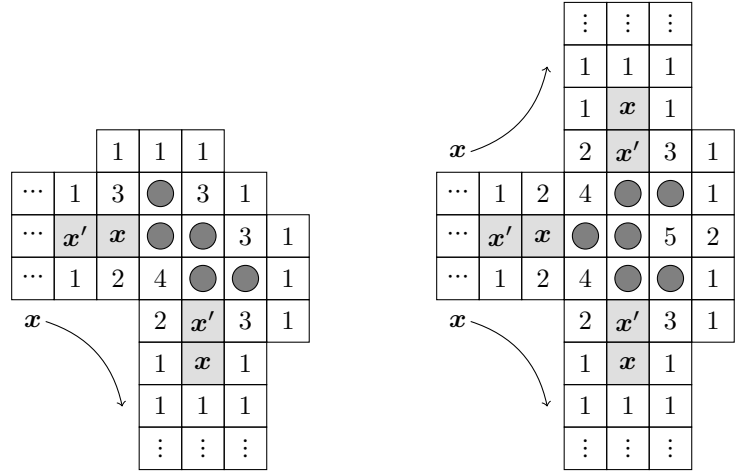

We also need to bend wires and to make a splitter to duplicate wires as Figure~\ref{fig:classicbs}. From now on, the gray circles in the nodes indicate mines those have been identified.

%%%%%%%%%%%%%%%%%%%%%%%%%%%%%%%%%%%%%%%%
% Figure 3 : classical bent & splitter %
%%%%%%%%%%%%%%%%%%%%%%%%%%%%%%%%%%%%%%%%

Using these basic components, existing studies propose some large computational components such as NOT, AND and OR gate. NOT and AND gates are created by Richard Kaye \cite{kaye00}, and the OR gate is made by Stefan \cite{kaye07}. %Some minesweeper configuration

\newpage

%%%%%%%%%%%%%%%%%%%%%%%%%%%%%%%%%
% Figure 4 : classical NOT gate %
%%%%%%%%%%%%%%%%%%%%%%%%%%%%%%%%%

\begin{figure}[ht]
\centering
\begin{tikzpicture} [rect/.style= {shape=regular polygon,regular polygon sides=4,minimum size=0.8cm, draw,inner sep=0,anchor=south}]

% fill in some colors

\foreach \k in {1,2,4,5,7,9,10,12,13} {\node [shape=regular polygon,regular polygon sides=4,minimum size=0.8cm,fill=lightgray!50!white,inner sep=0,anchor=south] at ({(\k)*0.565},{2*0.565}) {};} 

% generating nodes

\foreach \j in {0,4}{ 
  \foreach \i in {6,7,8}{
  \node[rect] (h\i;\j) at ({(\i)*0.565},{\j*0.565}) {};}  }      

\foreach \j in {1,...,3}{
  \foreach \i in {0,...,14}{
  \node[rect] (h\i;\j) at ({(\i)*0.565},{\j*0.565}) {};}  }      

% putting in contents

\foreach \k in {6,7,8} {
  \foreach \j in {0,4}
  {\node at (h\k;\j) {1};} }
\foreach \k in {1,2,3,4,5,9,10,11,12,13} {
  \foreach \j in {1,3}
  {\node at (h\k;\j) {1};} }
\foreach \k in {6,8} {
  \foreach \j in {1,3}
  {\node at (h\k;\j) {2};} }
\foreach \k in {7} {
  \foreach \j in {1,3}
  {\node [circle,draw,fill=gray,inner sep=3.5] at (h\k;\j) {};} }
  
\foreach \k in {3,11} {\node at (h\k;2) {1};}
\foreach \k in {6,8} {\node at (h\k;2) {3};}

\foreach \k in {2,5,9,12} {\node at (h\k;2) {$\boldsymbol{x}$};}
\foreach \k in {1,4,7,10,13} {\node at (h\k;2) {$\boldsymbol{x'}$};}

\foreach \k in {1,...,3}  {\node at (h0;\k) {$\cdots$};}
\foreach \k in {1,...,3}  {\node at (h14;\k) {$\cdots$};}

% arrows

{\node at ({2*0.565},{4.5*0.565}) {$\boldsymbol{x}\xrightarrow{\hspace*{1cm}}$};}
{\node at ({12*0.565},{4.5*0.565}) {$\boldsymbol{x'}\xrightarrow{\hspace*{1cm}}$};}

\end{tikzpicture}
\caption{A NOT gate on a normal Minesweeper}\label{classicnot}
\end{figure}

\vskip 2em

%%%%%%%%%%%%%%%%%%%%%%%%%%%%%%%%%
% Figure 5 : classical AND gate %
%%%%%%%%%%%%%%%%%%%%%%%%%%%%%%%%%

\begin{figure}[ht]
\centering
\begin{tikzpicture} [rect/.style= {shape=regular polygon,regular polygon sides=4,minimum size=0.8cm, draw,inner sep=0,anchor=south}]

% fill in some colors

\foreach \k in {2} {
  \foreach \j in {2,12}
  {\node [shape=regular polygon,regular polygon sides=4,minimum size=0.8cm,fill=lightgray!50!white,inner sep=0,anchor=south] at ({(\k)*0.565},{\j*0.565}) {};} }

\foreach \k in {2,8,12,14,15} {
  \foreach \j in {3,11}
  {\node [shape=regular polygon,regular polygon sides=4,minimum size=0.8cm,fill=lightgray!50!white,inner sep=0,anchor=south] at ({(\k)*0.565},{\j*0.565}) {};} }
  
\foreach \k in {9,...,11} {
  \foreach \j in {3,11}
  {\node [shape=regular polygon,regular polygon sides=4,minimum size=0.8cm,fill=RoyalBlue!35!white,inner sep=0,anchor=south] at ({(\k)*0.565},{\j*0.565}) {};} }
  
\foreach \k in {17} {
  \foreach \j in {4,10}
  {\node [shape=regular polygon,regular polygon sides=4,minimum size=0.8cm,fill=lightgray!50!white,inner sep=0,anchor=south] at ({(\k)*0.565},{\j*0.565}) {};} }
  
\foreach \k in {2,6} {
  \foreach \j in {5,9}
  {\node [shape=regular polygon,regular polygon sides=4,minimum size=0.8cm,fill=lightgray!50!white,inner sep=0,anchor=south] at ({(\k)*0.565},{\j*0.565}) {};} }
  
\foreach \k in {4,5,6,17} {
  \foreach \j in {6,8}
  {\node [shape=regular polygon,regular polygon sides=4,minimum size=0.8cm,fill=lightgray!50!white,inner sep=0,anchor=south] at ({(\k)*0.565},{\j*0.565}) {};} }
  
\foreach \k in {7,8,10,11,13,14,16,18,20,21} {
  \foreach \j in {7}
  {\node [shape=regular polygon,regular polygon sides=4,minimum size=0.8cm,fill=lightgray!50!white,inner sep=0,anchor=south] at ({(\k)*0.565},{\j*0.565}) {};} }
  
\foreach \k in {6} {
  \foreach \j in {7}
  {\node [shape=regular polygon,regular polygon sides=4,minimum size=0.8cm,fill=RoyalBlue!35!white,inner sep=0,anchor=south] at ({(\k)*0.565},{\j*0.565}) {$\boldsymbol{4}$};} }

% generating nodes

\foreach \j in {0,14}{ 
  \foreach \i in {1,2,3}{
  \node[rect] (h\i;\j) at ({(\i)*0.565},{\j*0.565}) {};}  }
  
\foreach \j in {1,13}{ 
  \foreach \i in {1,2,3,6,7,8,9,11,12,13,15,16,17}{
  \node[rect] (h\i;\j) at ({(\i)*0.565},{\j*0.565}) {};}  }          

\foreach \j in {2,12}{ 
  \foreach \i in {1,2,3,6,7,8,9,10,11,12,13,14,15,16,17,18,19}{
  \node[rect] (h\i;\j) at ({(\i)*0.565},{\j*0.565}) {};}  }
  
\foreach \j in {3,11}{
  \foreach \i in {1,...,19}{
  \node[rect] (h\i;\j) at ({(\i)*0.565},{\j*0.565}) {};}  }
  
\foreach \j in {4,5,9,10}{
  \foreach \i in {0,...,19}{
  \node[rect] (h\i;\j) at ({(\i)*0.565},{\j*0.565}) {};}  }
  
\foreach \j in {6,7,8}{ 
  \foreach \i in {0,...,23}{
  \node[rect] (h\i;\j) at ({(\i)*0.565},{\j*0.565}) {};}  }    
  
% Putting contents in the nodes

\foreach \k in {1,2,3} {
  \foreach \j in {0,14}  {\node at (h\k;\j) {$\vdots$};} }
\foreach \k in {23} {
  \foreach \j in {6,7,8}  {\node at (h\k;\j) {$\cdots$};} }
  
\foreach \k in {1,2,3,6,9,11,12,13,15,16,17} {
  \foreach \j in {1,13}  {\node at (h\k;\j) {1};} }
\foreach \k in {7,8} {
  \foreach \j in {1,13}  {\node at (h\k;\j) {2};} }
  
\foreach \k in {1,3,14,19} {
  \foreach \j in {2,12}  {\node at (h\k;\j) {1};} }
\foreach \k in {6,10,13,15,18} {
  \foreach \j in {2,12}  {\node at (h\k;\j) {2};} }
\foreach \k in {9,11,17} {
  \foreach \j in {2,12}  {\node at (h\k;\j) {3};} }
\foreach \k in {7,8,12,16} {
  \foreach \j in {2,12}  {\node [circle,draw,fill=gray,inner sep=3.5] at (h\k;\j) {};} }
{\node at (h2;2) {$\boldsymbol{v'}$};}
{\node at (h2;12) {$\boldsymbol{u'}$};}
  
\foreach \k in {1,3,4} {
  \foreach \j in {3,11}  {\node at (h\k;\j) {1};} }
\foreach \k in {5,19} {
  \foreach \j in {3,11}  {\node at (h\k;\j) {2};} }
\foreach \k in {13,16} {
  \foreach \j in {3,11}  {\node at (h\k;\j) {3};} }
\foreach \k in {6} {
  \foreach \j in {3,11}  {\node at (h\k;\j) {4};} }
\foreach \k in {12,15} {
  \foreach \j in {3,11}  {\node at (h\k;\j) {$\boldsymbol{t'}$};} }
\foreach \k in {14} {
  \foreach \j in {3,11}  {\node at (h\k;\j) {$\boldsymbol{t}$};} }
\foreach \k in {7,17,18} {
  \foreach \j in {3,11}  {\node [circle,draw,fill=gray,inner sep=3.5] at (h\k;\j) {};} }
{\node at (h8;3) {$\boldsymbol{r}$};}
{\node at (h2;3) {$\boldsymbol{v}$};}
\foreach \k in {9,10,11} {
  \foreach \j in {3} {
  \pgfmathtruncatemacro\m{\k-8}  {\node at (h\k;\j) {$\boldsymbol{b_{\m}}$};} } }
{\node at (h8;11) {$\boldsymbol{s}$};}
{\node at (h2;11) {$\boldsymbol{u}$};}
\foreach \k in {9,10,11} {
  \foreach \j in {11} {
  \pgfmathtruncatemacro\m{\k-8}  {\node at (h\k;\j) {$\boldsymbol{a_{\m}}$};} } }

\foreach \k in {0,3,4,14,15} {
  \foreach \j in {4,10}  {\node at (h\k;\j) {1};} }
\foreach \k in {1,2,10,13,16,19} {
  \foreach \j in {4,10}  {\node at (h\k;\j) {2};} }
\foreach \k in {9,11} {
  \foreach \j in {4,10}  {\node at (h\k;\j) {3};} }
\foreach \k in {7} {
  \foreach \j in {4,10}  {\node at (h\k;\j) {4};} }
\foreach \k in {17} {
  \foreach \j in {4,10}  {\node at (h\k;\j) {$\boldsymbol{t}$};} }
\foreach \k in {5,6,8,12,18} {
  \foreach \j in {4,10}  {\node [circle,draw,fill=gray,inner sep=3.5] at (h\k;\j) {};} }

\foreach \k in {10,14,15} {
  \foreach \j in {5,9}  {\node at (h\k;\j) {0};} }
\foreach \k in {8,9,11,12,13,16,19} {
  \foreach \j in {5,9}  {\node at (h\k;\j) {1};} }
\foreach \k in {0,3,4,17,18} {
  \foreach \j in {5,9}  {\node at (h\k;\j) {2};} }
\foreach \k in {7} {
  \foreach \j in {5,9}  {\node at (h\k;\j) {3};} }
\foreach \k in {5} {
  \foreach \j in {5,9}  {\node at (h\k;\j) {4};} }
\foreach \k in {1} {
  \foreach \j in {5,9}  {\node [circle,draw,fill=gray,inner sep=3.5] at (h\k;\j) {};} }
{\node at (h2;5) {$\boldsymbol{v'}$};}
{\node at (h6;5) {$\boldsymbol{r'}$};}
{\node at (h2;9) {$\boldsymbol{u'}$};}
{\node at (h6;9) {$\boldsymbol{s'}$};}
  
\foreach \k in {8,...,16} {
  \foreach \j in {6,8}  {\node at (h\k;\j) {1};} }
\foreach \k in {18,...,22} {
  \foreach \j in {6,8}  {\node at (h\k;\j) {1};} }
\foreach \k in {0,7} {
  \foreach \j in {6,8}  {\node at (h\k;\j) {2};} }
\foreach \k in {3} {
  \foreach \j in {6,8}  {\node at (h\k;\j) {3};} }
\foreach \k in {17} {
  \foreach \j in {6,8}  {\node at (h\k;\j) {$\boldsymbol{t'}$};} }
\foreach \k in {1,2} {
  \foreach \j in {6,8}  {\node [circle,draw,fill=gray,inner sep=3.5] at (h\k;\j) {};} }
{\node at (h4;6) {$\boldsymbol{v}$};}
{\node at (h5;6) {$\boldsymbol{v'}$};}
{\node at (h6;6) {$\boldsymbol{r}$};}
{\node at (h4;8) {$\boldsymbol{u}$};}
{\node at (h5;8) {$\boldsymbol{u'}$};}
{\node at (h6;8) {$\boldsymbol{s}$};}
  
\foreach \k in {8,11,14,20} {\node at (h\k;7) {$\boldsymbol{t'}$};}
\foreach \k in {7,10,13,16,18,21} {\node at (h\k;7) {$\boldsymbol{t}$};}
\foreach \k in {3,5} {\node [circle,draw,fill=gray,inner sep=3.5] at (h\k;7) {};}
\foreach \k in {9,12,15,19,22} {\node at (h\k;7) {1};}
\foreach \k in {0,17} {\node at (h\k;7) {2};}
\foreach \k in {1,4} {\node at (h\k;7) {4};}
\foreach \k in {2} {\node at (h\k;7) {5};}

% arrows

{\node at ({0*0.565},{14*0.565}) {$\boldsymbol{u}$};}
{\node at ({0*0.565},{12.5*0.565}) {\begin{turn}{-90}$\xrightarrow{\hspace*{1cm}}$\end{turn}};}

{\node at ({0*0.565},{1*0.565}) {$\boldsymbol{v}$};}
{\node at ({0*0.565},{2.5*0.565}) {\begin{turn}{90}$\xrightarrow{\hspace*{1cm}}$\end{turn}};}

{\node at ({21.5*0.565},{9.5*0.565}) {$\boldsymbol{t}\xrightarrow{\hspace*{1cm}}$};}

\end{tikzpicture}
\caption{An AND gate on a normal Minesweeper}\label{classicand}
\end{figure}

\vskip 2em

The reason that Figure~\ref{classicand} is actually an AND gate is as follows. If the result is T, then $\boldsymbol{t}$ must be T, which means that $\boldsymbol{a_2,\,a_3}$ are both T. Then $\boldsymbol{a_1}$ should be F, so $\boldsymbol{s}$ must be T, and by the symmetry of this logic gate, $\boldsymbol{r}$ also should be T. It implies that $\boldsymbol{u'}$ and $\boldsymbol{v'}$ are both F when we observe the blue square named `$\boldsymbol{4}$'. Therefore, the result($\boldsymbol{r}$) is T only when two inputs($\boldsymbol{u}$ and $\boldsymbol{v}$) are both T, which concludes that this logic gate is actually an AND gate. The truth table for this AND gate is shown in table~\ref{tab:table2}:

\newpage

\begin{table}
\centering
\caption{Truth table for the AND gate}
\vskip 0.3em
\begin{tabular}{|c|c||c|c|c|c|c|c|c|c||c|}
\hline
$\boldsymbol{u}$ & $\boldsymbol{v}$ & $\boldsymbol{s}$ & $\boldsymbol{r}$ & $\boldsymbol{a_1}$ & $\boldsymbol{a_2}$ & $\boldsymbol{a_3}$ & $\boldsymbol{b_1}$ & $\boldsymbol{b_2}$ & $\boldsymbol{b_3}$ & $\boldsymbol{t}=\boldsymbol{u}\wedge\boldsymbol{v}$\\\hline\hline
T&T&T&T&F&T&T&F&T&T&T\\\hline
T&F&T&T&T&F&T&T&F&T&F\\\hline
F&T&T&T&T&F&T&T&F&T&F\\\hline
F&F&F&F&T&T&F&T&T&F&F\\\hline
\end{tabular}
\label{tab:table2}
\end{table}

Next figure is an OR gate made by Stefan \cite{kaye07}. Similarly we can easily check that the result($r$) is F only when two inputs($u$ and $v$) are both F.

\vskip 1em

%%%%%%%%%%%%%%%%%%%%%%%%%%%%%%%%
% Figure 6 : classical OR gate %
%%%%%%%%%%%%%%%%%%%%%%%%%%%%%%%%

\begin{figure}[ht]
\centering
\begin{tikzpicture} [rect/.style= {shape=regular polygon,regular polygon sides=4,minimum size=0.8cm, draw,inner sep=0,anchor=south}]

% fill in some colors

\foreach \k in {6,10} {
  \foreach \j in {2}
  {\node [shape=regular polygon,regular polygon sides=4,minimum size=0.8cm,fill=lightgray!50!white,inner sep=0,anchor=south] at ({(\k)*0.565},{\j*0.565}) {};} }

\foreach \k in {6,11} {
  \foreach \j in {4,6}
  {\node [shape=regular polygon,regular polygon sides=4,minimum size=0.8cm,fill=lightgray!50!white,inner sep=0,anchor=south] at ({(\k)*0.565},{\j*0.565}) {};} }
  
\foreach \k in {7,8,9} {
  \foreach \j in {2}
  {\node [shape=regular polygon,regular polygon sides=4,minimum size=0.8cm,fill=RoyalBlue!35!white,inner sep=0,anchor=south] at ({(\k)*0.565},{\j*0.565}) {};} }
  
\foreach \k in {2,3,5,7,8,10,12,14,15} {
  \foreach \j in {5}
  {\node [shape=regular polygon,regular polygon sides=4,minimum size=0.8cm,fill=lightgray!50!white,inner sep=0,anchor=south] at ({(\k)*0.565},{\j*0.565}) {};} }
  
\foreach \k in {6} {
  \foreach \j in {5}
  {\node [shape=regular polygon,regular polygon sides=4,minimum size=0.8cm,fill=RoyalBlue!35!white,inner sep=0,anchor=south] at ({(\k)*0.565},{\j*0.565}) {};} }
  
\foreach \k in {6} {
  \foreach \j in {8,9}
  {\node [shape=regular polygon,regular polygon sides=4,minimum size=0.8cm,fill=lightgray!50!white,inner sep=0,anchor=south] at ({(\k)*0.565},{\j*0.565}) {};} }

% generating nodes

\foreach \j in {0}{
  \foreach \i in {4,5,6,7,9,10,11,12}{
  \node[rect] (h\i;\j) at ({(\i)*0.565},{\j*0.565}) {};}  }
  
\foreach \j in {1}{ 
  \foreach \i in {4,...,12}{
  \node[rect] (h\i;\j) at ({(\i)*0.565},{\j*0.565}) {};}  }
  
\foreach \j in {2,3,7}{ 
  \foreach \i in {4,...,13}{
  \node[rect] (h\i;\j) at ({(\i)*0.565},{\j*0.565}) {};}  }
  
\foreach \j in {4,5,6}{ 
  \foreach \i in {0,...,17}{
  \node[rect] (h\i;\j) at ({(\i)*0.565},{\j*0.565}) {};}  }
  
\foreach \j in {8}{
  \foreach \i in {5,6,7,9,10,11,12,13}{
  \node[rect] (h\i;\j) at ({(\i)*0.565},{\j*0.565}) {};}  }
  
\foreach \j in {9,10,11}{
  \foreach \i in {5,6,7}{
  \node[rect] (h\i;\j) at ({(\i)*0.565},{\j*0.565}) {};}  }

% Putting contents in the nodes

\foreach \k in {5,6,7} {
  \foreach \j in {11}
  {\node at (h\k;\j) {$\vdots$};} }
\foreach \k in {0,17} {
  \foreach \j in {4,5,6}
  {\node at (h\k;\j) {$\cdots$};} }
  
\foreach \k in {4,7,9,12} {
  \foreach \j in {0}
  {\node at (h\k;\j) {1};} }
\foreach \k in {5,6,10,11} {
  \foreach \j in {0}
  {\node at (h\k;\j) {2};} }
 
\foreach \k in {5,6,10,11} {
  \foreach \j in {1}
  {\node [circle,draw,fill=gray,inner sep=3.5] at (h\k;\j) {};} }
\foreach \k in {4,8,12} {
  \foreach \j in {1}
  {\node at (h\k;\j) {2};} }
\foreach \k in {7,9} {
  \foreach \j in {1}
  {\node at (h\k;\j) {3};} }
  
\foreach \k in {7,8,9} {
  \foreach \j in {2} {
  \pgfmathtruncatemacro\m{\k-6}
  {\node at (h\k;\j) {$\boldsymbol{a_{\m}}$};} } }
\foreach \k in {13} {
  \foreach \j in {2}
  {\node at (h\k;\j) {1};} }
\foreach \k in {4} {
  \foreach \j in {2}
  {\node at (h\k;\j) {2};} }
\foreach \k in {12} {
  \foreach \j in {2}
  {\node at (h\k;\j) {3};} }
\foreach \k in {6} {
  \foreach \j in {2}
  {\node at (h\k;\j) {$\boldsymbol{s'}$};} }
\foreach \k in {10} {
  \foreach \j in {2}
  {\node at (h\k;\j) {$\boldsymbol{r}$};} }
\foreach \k in {5,11} {
  \foreach \j in {2}
  {\node [circle,draw,fill=gray,inner sep=3.5] at (h\k;\j) {};} }
  
\foreach \k in {6,7,8,9,10,12} {
  \foreach \j in {3}
  {\node [circle,draw,fill=gray,inner sep=3.5] at (h\k;\j) {};} }
\foreach \k in {13} {
  \foreach \j in {3}
  {\node at (h\k;\j) {1};} }
\foreach \k in {4} {
  \foreach \j in {3}
  {\node at (h\k;\j) {2};} }
\foreach \k in {5,11} {
  \foreach \j in {3}
  {\node at (h\k;\j) {4};} }

\foreach \k in {1,2,3,14,15,16} {
  \foreach \j in {4,6}
  {\node at (h\k;\j) {1};} }
\foreach \k in {4,13} {
  \foreach \j in {4,6}
  {\node at (h\k;\j) {2};} }
\foreach \k in {10} {
  \foreach \j in {4,6}
  {\node at (h\k;\j) {3};} }
\foreach \k in {11} {
  \foreach \j in {4,6}
  {\node at (h\k;\j) {$\boldsymbol{r'}$};} }
\foreach \k in {5,7} {
  \foreach \j in {4,6}
  {\node [circle,draw,fill=gray,inner sep=3.5] at (h\k;\j) {};} }

\foreach \k in {6} {
  \foreach \j in {4}
  {\node at (h\k;\j) {$\boldsymbol{s}$};} }
\foreach \k in {8} {
  \foreach \j in {4}
  {\node at (h\k;\j) {5};} }
\foreach \k in {9} {
  \foreach \j in {4}
  {\node at (h\k;\j) {4};} }
\foreach \k in {12} {
  \foreach \j in {4}
  {\node at (h\k;\j) {2};} }
  
\foreach \k in {6} {
  \foreach \j in {6}
  {\node at (h\k;\j) {$\boldsymbol{u'}$};} }
\foreach \k in {8,9} {
  \foreach \j in {6}
  {\node at (h\k;\j) {2};} }
\foreach \k in {12} {
  \foreach \j in {6}
  {\node at (h\k;\j) {3};} }
  
\foreach \k in {1,9,13,16} {
  \foreach \j in {5}
  {\node at (h\k;\j) {1};} }
\foreach \k in {11} {
  \foreach \j in {5}
  {\node at (h\k;\j) {2};} }
\foreach \k in {4} {
  \foreach \j in {5}
  {\node at (h\k;\j) {3};} }
\foreach \k in {6} {
  \foreach \j in {5}
  {\node at (h\k;\j) {{\textbf{6}}};} }
\foreach \k in {2,5} {
  \foreach \j in {5}
  {\node at (h\k;\j) {$\boldsymbol{v'}$};} }
\foreach \k in {3} {
  \foreach \j in {5}
  {\node at (h\k;\j) {$\boldsymbol{v}$};} }
\foreach \k in {7,10,12,15} {
  \foreach \j in {5}
  {\node at (h\k;\j) {$\boldsymbol{r}$};} }
\foreach \k in {8,14} {
  \foreach \j in {5}
  {\node at (h\k;\j) {$\boldsymbol{r'}$};} }
  
\foreach \k in {4,8,9,13} {
  \foreach \j in {7}
  {\node at (h\k;\j) {1};} }
\foreach \k in {5,7} {
  \foreach \j in {7}
  {\node at (h\k;\j) {2};} }
\foreach \k in {6} {
  \foreach \j in {7}
  {\node at (h\k;\j) {3};} }
\foreach \k in {10,11,12} {
  \foreach \j in {7}
  {\node [circle,draw,fill=gray,inner sep=3.5] at (h\k;\j) {};} }
  
\foreach \k in {5,7,9,13} {
  \foreach \j in {8}
  {\node at (h\k;\j) {1};} }
\foreach \k in {10,12} {
  \foreach \j in {8}
  {\node at (h\k;\j) {2};} }
\foreach \k in {11} {
  \foreach \j in {8}
  {\node at (h\k;\j) {3};} }
\foreach \k in {6} {
  \foreach \j in {8}
  {\node at (h\k;\j) {$\boldsymbol{u}$};} }
  
\foreach \k in {6} {
  \foreach \j in {9}
  {\node at (h\k;\j) {$\boldsymbol{u'}$};} }
\foreach \k in {5,7} {
  \foreach \j in {9,10}
  {\node at (h\k;\j) {1};} }
\foreach \k in {6} {
  \foreach \j in {10}
  {\node at (h\k;\j) {1};} }
  
{\node at ({4*0.565},{11*0.565}) {$\boldsymbol{u}$};}
{\node at ({4*0.565},{9.5*0.565}) {\begin{turn}{-90}$\xrightarrow{\hspace*{1cm}}$\end{turn}};}

{\node at ({1.5*0.565},{3.5*0.565}) {$\boldsymbol{v}\xrightarrow{\hspace*{1cm}}$};}
{\node at ({15.5*0.565},{3.5*0.565}) {$\boldsymbol{r}\xrightarrow{\hspace*{1cm}}$};}

\end{tikzpicture}
\caption{An OR gate on a normal Minesweeper}\label{classicor}
\end{figure}
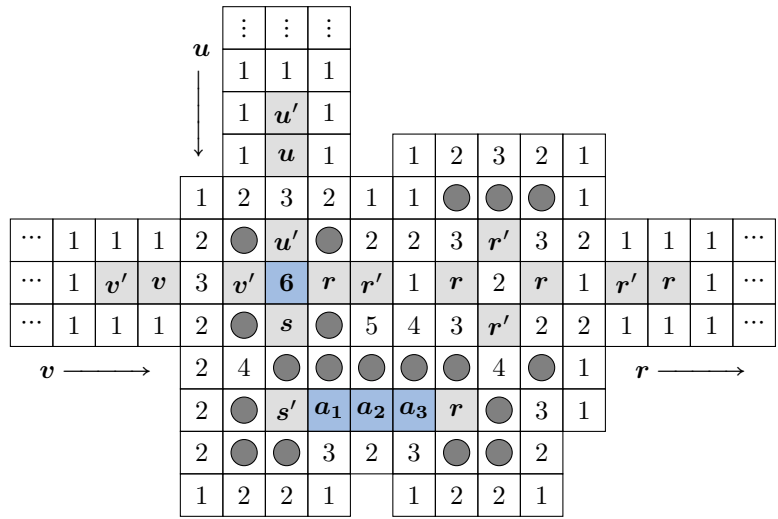

\vskip 1em

The truth table for this OR gate is as below:

\begin{table}[!htbp]
\centering
\caption{Truth table for the OR gate}
\vskip 0.3em
\begin{tabular}{|c|c||c|c|c|c||c|}
\hline
$\boldsymbol{u}$ & $\boldsymbol{v}$ & $\boldsymbol{s}$ & $\boldsymbol{a_1}$ & $\boldsymbol{a_2}$ & $\boldsymbol{a_3}$ & $\boldsymbol{r}=\boldsymbol{u}\vee\boldsymbol{v}$\\\hline\hline
T&T&T&T&T&F&T\\\hline
T&F&F&T&F&T&T\\\hline
F&T&F&T&F&T&T\\\hline
F&F&F&F&T&T&F\\\hline
\end{tabular}
\label{tab:table4}
\end{table}

\newpage

%%%%%%%%%%%%%%%%%%%%%%%%%%%%%%%%%%%%%%%%%%%%%%%%%%%%%%%%%%%%%%%%%%
%%%%% 3. Computational components in a hexagonal Minesweeper %%%%%
%%%%%%%%%%%%%%%%%%%%%%%%%%%%%%%%%%%%%%%%%%%%%%%%%%%%%%%%%%%%%%%%%%

\section{Computational components in a hexagonal\\Minesweeper}
\label{sec3}

As I discussed in the introduction, we can apply the computational components on a hexagonal grid as Bondt \cite{bondt12} made a hexagonal wire.

%%%%%%%%%%%%%%%%%%%%%%%%%%%%%%%%%%%%%%%%%%%%%%%%%%%%%%%%
% Figure 7 : (original) hexagonal wire and a NOT gate? %
%%%%%%%%%%%%%%%%%%%%%%%%%%%%%%%%%%%%%%%%%%%%%%%%%%%%%%%%

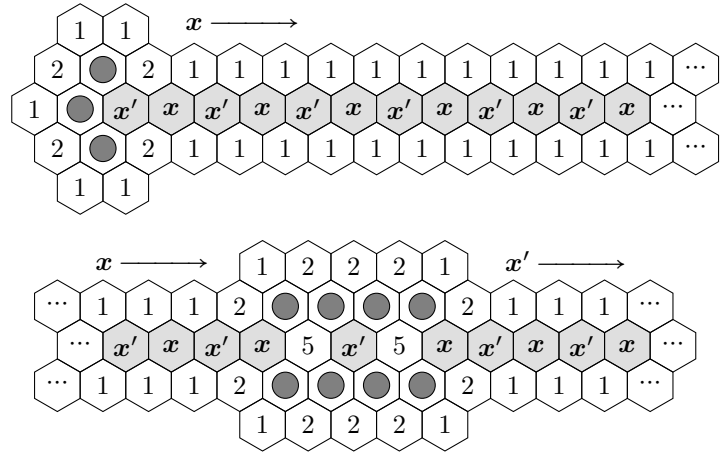
\begin{figure}[ht]
\centering
\begin{tikzpicture} [hexa/.style= {shape=regular polygon,regular polygon sides=6,minimum size=0.7cm, draw,inner sep=0,anchor=south,rotate=30}]

% fill in some colors

\foreach \k in {2,...,13} {\node [shape=regular polygon,regular polygon sides=6,minimum size=0.7cm,fill=lightgray!50!white,inner sep=0,anchor=south,rotate=30] at ({(\k-2/2)*sin(60)*0.7},{2*0.75*0.7}) {};}
\foreach \k in {-1,0,1,2,4,6,7,8,9,10} {\node [shape=regular polygon,regular polygon sides=6,minimum size=0.7cm,fill=lightgray!50!white,inner sep=0,anchor=south,rotate=30] at ({(\k-(-4)/2)*sin(60)*0.7},{-4*0.75*0.7}) {};}

% generating nodes

\foreach \i in {0,1}{
  \node[hexa] (h\i;0) at ({(\i-0/2)*sin(60)*0.7},{0*0.75*0.7}) {};}
\foreach \j in {1,...,2}{
  \foreach \i in {0,...,14}{
  \node[hexa] (h\i;\j) at ({(\i-\j/2)*sin(60)*0.7},{\j*0.75*0.7}) {};}  }      
\foreach \i in {0,...,14}{ 
  \node[hexa] (h\i;3) at ({(\i+3/2-2)*sin(60)*0.7},{3*0.75*0.7}) {};}
\foreach \i in {0,1}{ 
  \node[hexa] (h\i;4) at ({(\i+4/2-2)*sin(60)*0.7},{4*0.75*0.7}) {};}
\foreach \i in {3,...,7}{ 
  \node[hexa] (h\i;-2) at ({(\i-(-2)/2)*sin(60)*0.7},{-2*0.75*0.7}) {};}
\foreach \i in {-2,...,11}{ 
  \node[hexa] (h\i;-3) at ({((\i-(-3)/2)*sin(60)*0.7},{-3*0.75*0.7}) {};}
\foreach \i in {-2,...,11}{ 
  \node[hexa] (h\i;-4) at ({((\i-(-4)/2)*sin(60)*0.7},{-4*0.75*0.7}) {};}
\foreach \i in {-3,...,10}{ 
  \node[hexa] (h\i;-5) at ({((\i-(-5)/2)*sin(60)*0.7},{-5*0.75*0.7}) {};}
\foreach \i in {1,...,5}{ 
  \node[hexa] (h\i;-6) at ({((\i-(-6)/2)*sin(60)*0.7},{-6*0.75*0.7}) {};}
  
% putting contents in the nodes

\foreach \k in {1,2,3}  {\node at (h14;\k) {$\cdots$};}

\foreach \k in {1,2,3} {\node [circle,draw,fill=gray,inner sep=3.5] at (h1;\k) {};}

\foreach \k in {1,3} {
  \foreach \j in {3,...,13}  {\node at (h\j;\k) {1};} }
\foreach \k in {1,3} {
  \foreach \j in {0,2}  {\node at (h\j;\k) {2};} }
\foreach \k in {0,4} {
  \foreach \j in {0,1}  {\node at (h\j;\k) {1};} }
  
\foreach \k in {3,7}  {\node at (h\k;-2) {1};}
\foreach \k in {4,5,6}  {\node at (h\k;-2) {2};}

\foreach \k in {-2,11} {
  \foreach \j in {-3,-4}  {\node at (h\k;\j) {$\cdots$};} }
\foreach \k in {-1,0,1,8,9,10}  {\node at (h\k;-3) {1};}
\foreach \k in {2,7}  {\node at (h\k;-3) {2};}
\foreach \k in {3,...,6} {\node [circle,draw,fill=gray,inner sep=3.5] at (h\k;-3) {};}

\foreach \k in {3,5}  {\node at (h\k;-4) {5};}
\foreach \k in {-1,1,4,7,9}  {\node at (h\k;-4) {$\boldsymbol{x'}$};}
\foreach \k in {0,2,6,8,10}  {\node at (h\k;-4) {$\boldsymbol{x}$};}

\foreach \k in {-3,10}  {\node at (h\k;-5) {$\cdots$};}
\foreach \k in {-2,-1,0,7,8,9}  {\node at (h\k;-5) {1};}
\foreach \k in {1,6}  {\node at (h\k;-5) {2};}
\foreach \k in {2,...,5} {\node [circle,draw,fill=gray,inner sep=3.5] at (h\k;-5) {};}

\foreach \k in {1,5}  {\node at (h\k;-6) {1};}
\foreach \k in {2,3,4}  {\node at (h\k;-6) {2};}

{\node at (h0;2) {1};}
\foreach \k in {2,4,6,8,10,12}  {\node at (h\k;2) {$\boldsymbol{x'}$};}
\foreach \k in {3,5,7,9,11,13}  {\node at (h\k;2) {$\boldsymbol{x}$};}

% arrows

{\node at (2,2.5) {$\boldsymbol{x}\xrightarrow{\hspace*{1cm}}$};}
{\node at (0.8,-0.7) {$\boldsymbol{x}\xrightarrow{\hspace*{1cm}}$};}
{\node at (6.3,-0.7) {$\boldsymbol{x'}\xrightarrow{\hspace*{1cm}}$};}

\end{tikzpicture}
\caption{Hexagonal wire and a NOT gate \cite{bondt12}}\label{original}
\end{figure}

\begin{figure}[ht]
\centering
$\boldsymbol{x} \xrightarrow{\hspace*{2cm}}$
\begin{tikzpicture} [hexa/.style= {shape=regular polygon,regular polygon sides=6,minimum size=0.7cm, draw,inner sep=0,anchor=south,rotate=30}]

% fill in some colors

\foreach \k in {2,3,5,6,8,9,11,12} {\node [shape=regular polygon,regular polygon sides=6,minimum size=0.7cm,fill=lightgray!50!white,inner sep=0,anchor=south,rotate=30] at ({(\k-2/2)*sin(60)*0.7},{2*0.75*0.7}) {};} 

% generating nodes

\foreach \j in {0,...,2}{ 
  \foreach \i in {0,...,14}{
  \node[hexa] (h\i;\j) at ({(\i-\j/2)*sin(60)*0.7},{\j*0.75*0.7}) {};}  }      
\foreach \j in {3,4}{ 
  \foreach \i in {0,...,14}{
  \node[hexa] (h\i;\j) at ({(\i+\j/2-2)*sin(60)*0.7},{\j*0.75*0.7}) {};}  } 
  
% putting in contents

 \foreach \k in {0,2,4}  {\node at (h0;\k) {1};} 
 \foreach \k in {1,3}  {\node at (h0;\k) {2};} 
 \foreach \k in {0,4} {\node at (h1;\k) {1};} 
 \foreach \k in {1,2,3} {\node [circle,draw,fill=gray,inner sep=3.5] at (h1;\k) {};} 

\foreach \k in {2,5,8,11}  {\node at (h\k;2) {$\boldsymbol{x'}$};}
\foreach \k in {3,6,9,12}  {\node at (h\k;2) {$\boldsymbol{x}$};}
\foreach \k in {4,7,10,13}  {\node at (h\k;2) {5};}
\foreach \k in {2,5,8,11}  {\node at (h\k;1) {3};}
\foreach \k in {2,5,8,11}  {\node at (h\k;3) {3};}
\foreach \k in {3,4,6,7,9,10,12,13}  {\node [circle,draw,fill=gray,inner sep=3.5] at (h\k;1) {};}
\foreach \k in {3,4,6,7,9,10,12,13}  {\node [circle,draw,fill=gray,inner sep=3.5] at (h\k;3) {};}

\foreach \k in {2,4,5,7,8,10,11,13}  {\node at (h\k;0) {1};}
\foreach \k in {2,4,5,7,8,10,11,13}  {\node at (h\k;4) {1};}
\foreach \k in {3,6,9,12}  {\node at (h\k;0) {2};}
\foreach \k in {3,6,9,12}  {\node at (h\k;4) {2};}
\foreach \k in {0,...,4}  {\node at (h14;\k) {$\cdots$};}

\end{tikzpicture}
\caption{An improved form of a wire on a hexagonal Minesweeper}\label{wire}
\end{figure}
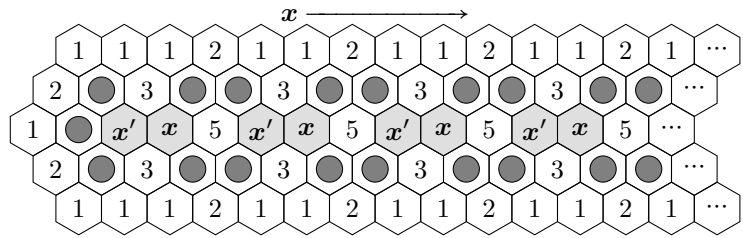

Even though this application is only one of the possible forms of wires, the fact that we cannot distinguish `0' and `1' in an infinite wire without a starting point, supports the need for an improved form of a hexagonal wire \cite{bondt12}. The figure~\ref{wire} below represents an improved form of wire on a hexagonal Minesweeper.

%%%%%%%%%%%%%%%%%%%%%%%%%%%%%
% Figure 8 : hexagonal wire %
%%%%%%%%%%%%%%%%%%%%%%%%%%%%%

This form of wire allows us to distinguish $x$ and $x'$ clearly; in other words, it represents a phase of Boolean values. A NOT gate using this wire perfectly demonstrate \textit{false} and \textit{true} respectively. Of course, we can make a curve and a splitter of wires.

\newpage

%%%%%%%%%%%%%%%%%%%%%%%%%%%%%%%%
% Figure 9 : A curve of a wire %
%%%%%%%%%%%%%%%%%%%%%%%%%%%%%%%%

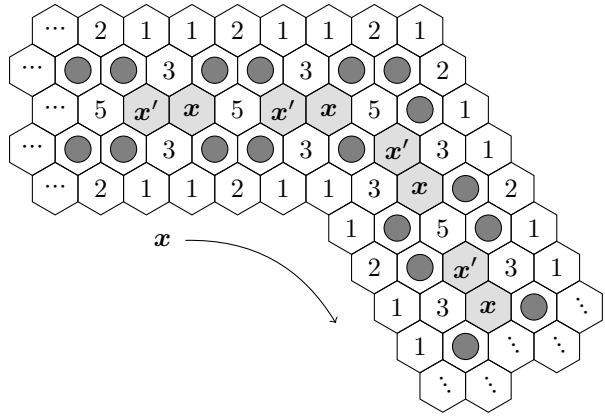
\begin{figure}[ht]
\centering
\begin{tikzpicture} [hexa/.style= {shape=regular polygon,regular polygon sides=6,minimum size=0.7cm, draw,inner sep=0,anchor=south,rotate=30}]

% Fill in some color

\foreach \k in {2,3,5,6} {\node [shape=regular polygon,regular polygon sides=6,minimum size=0.7cm,fill=lightgray!50!white,inner sep=0,anchor=south,rotate=30] at ({(\k-2/2)*sin(60)*0.7},{2*0.75*0.7}) {};}
\foreach \k in {1,0,-2,-3} {\node [shape=regular polygon,regular polygon sides=6,minimum size=0.7cm,fill=lightgray!50!white,inner sep=0,anchor=south,rotate=30] at ({(7-\k/2)*sin(60)*0.7},{\k*0.75*0.7}) {};}

% generating nodes

\foreach \j in {0,...,2}{
  \foreach \i in {0,...,9}{
  \node[hexa] (h\i;\j) at ({(\i-\j/2)*sin(60)*0.7},{\j*0.75*0.7}) {};}  }      
\foreach \j in {3,4}{
  \pgfmathsetmacro\end{11-\j} 
  \foreach \i in {0,...,\end}{
  \node[hexa] (h\i;\j) at ({(\i+\j/2-2)*sin(60)*0.7},{\j*0.75*0.7}) {};}  } 
  
\foreach \j in {0,1}{
  \node[hexa] (h-1;\j) at ({(-1-\j/2)*sin(60)*0.7},{\j*0.75*0.7}) {}; }
\foreach \j in {3,4}{
  \node[hexa] (h-1;\j) at ({(-1+(\j-3)/2-0.5)*sin(60)*0.7},{\j*0.75*0.7}) {}; } 
\foreach \j in {-3,...,-1}{
  \foreach \i in {5,...,9}{
  \node[hexa] (h\i;\j) at ({(\i-\j/2)*sin(60)*0.7},{\j*0.75*0.7}) {};}  } 
\foreach \i in {5,...,8}{
  \node[hexa] (h\i;-4) at ({(\i-(-4)/2)*sin(60)*0.7},{-4*0.75*0.7}) {};}  
\foreach \i in {5,6}{
  \node[hexa] (h\i;-5) at ({(\i-(-5)/2)*sin(60)*0.7},{-5*0.75*0.7}) {};}  

% Putting contents in the nodes

\foreach \k in {5,6}  {\node at (h\k;-5) {\begin{turn}{-60}$\cdots$\end{turn}};}

\foreach \k in {7,8}  {\node at (h\k;-4) {\begin{turn}{-60}$\cdots$\end{turn}};}
\foreach \k in {5}  {\node at (h\k;-4) {1};}
\foreach \k in {6} {\node [circle,draw,fill=gray,inner sep=3.5] at (h\k;-4) {};}

\foreach \k in {9}  {\node at (h\k;-3) {\begin{turn}{-60}$\cdots$\end{turn}};}
\foreach \k in {5}  {\node at (h\k;-3) {1};}
\foreach \k in {6}  {\node at (h\k;-3) {3};}
\foreach \k in {7}  {\node at (h\k;-3) {$\boldsymbol{x}$};}
\foreach \k in {8} {\node [circle,draw,fill=gray,inner sep=3.5] at (h\k;-3) {};}

\foreach \k in {9}  {\node at (h\k;-2) {1};}
\foreach \k in {5}  {\node at (h\k;-2) {2};}
\foreach \k in {8}  {\node at (h\k;-2) {3};}
\foreach \k in {7}  {\node at (h\k;-2) {$\boldsymbol{x'}$};}
\foreach \k in {6} {\node [circle,draw,fill=gray,inner sep=3.5] at (h\k;-2) {};}

\foreach \k in {5,9}  {\node at (h\k;-1) {1};}
\foreach \k in {7}  {\node at (h\k;-1) {5};}
\foreach \k in {6,8} {\node [circle,draw,fill=gray,inner sep=3.5] at (h\k;-1) {};}

\foreach \k in {-1} {
  \foreach \j in {0,1,3,4}
  {\node at (h\k;\j) {$\cdots$};} }
\foreach \k in {1,2,4,5} {
  \foreach \j in {0,4}
  {\node at (h\k;\j) {1};} }
\foreach \k in {0,3} {
  \foreach \j in {0,4}
  {\node at (h\k;\j) {2};} }
  
\foreach \k in {9}  {\node at (h\k;0) {2};}
\foreach \k in {6}  {\node at (h\k;0) {3};}
\foreach \k in {7}  {\node at (h\k;0) {$\boldsymbol{x}$};}
\foreach \k in {8} {\node [circle,draw,fill=gray,inner sep=3.5] at (h\k;0) {};}

\foreach \k in {6}  {\node at (h\k;4) {2};}
\foreach \k in {7}  {\node at (h\k;4) {1};}

\foreach \k in {0,1,3,4,6} {
  \foreach \j in {1,3}
  {\node [circle,draw,fill=gray,inner sep=3.5] at (h\k;\j) {};} }
\foreach \k in {2,5} {
  \foreach \j in {1,3}
  {\node at (h\k;\j) {3};} }
  
\foreach \k in {7}  {\node at (h\k;1) {$\boldsymbol{x'}$};}
\foreach \k in {8}  {\node at (h\k;1) {3};}
\foreach \k in {9}  {\node at (h\k;1) {1};}

\foreach \k in {8}  {\node at (h\k;3) {2};}
\foreach \k in {7} {\node [circle,draw,fill=gray,inner sep=3.5] at (h\k;3) {};}

\foreach \k in {0}  {\node at (h\k;2) {$\cdots$};}
\foreach \k in {9}  {\node at (h\k;2) {1};}
\foreach \k in {1,4,7}  {\node at (h\k;2) {5};}
\foreach \k in {2,5}  {\node at (h\k;2) {$\boldsymbol{x'}$};}
\foreach \k in {3,6}  {\node at (h\k;2) {$\boldsymbol{x}$};}
\foreach \k in {8} {\node [circle,draw,fill=gray,inner sep=3.5] at (h\k;2) {};}

% arrows

{\node at ({((0.7)-(-0.8)/2)*sin(60)*0.7},{-0.8*0.75*0.7}) {$\boldsymbol{x}$};}
\draw [->] ({((1.2)-(-0.8)/2)*sin(60)*0.7},{-0.8*0.75*0.7}) to [bend left] ({((1.2)-(-0.8)/2)*sin(60)*0.7+2},{-0.8*0.75*0.7-1.1});

\end{tikzpicture}
\caption{A curve of a wire}\label{curve}
\end{figure}

\vskip 3em

%%%%%%%%%%%%%%%%%%%%%%%%%%%%%%%%%%%%%%%%%%%%%%%%%
%  Figure 10 : A splitter by merging two curves %
%%%%%%%%%%%%%%%%%%%%%%%%%%%%%%%%%%%%%%%%%%%%%%%%%

\begin{figure}[!ht]
\centering
\begin{tikzpicture} [hexa/.style= {shape=regular polygon,regular polygon sides=6,minimum size=0.7cm, draw,inner sep=0,anchor=south,rotate=30}]

% Fill in some colors

\foreach \k in {2,3,5,6} {\node [shape=regular polygon,regular polygon sides=6,minimum size=0.7cm,fill=lightgray!50!white,inner sep=0,anchor=south,rotate=30] at ({(\k-2/2)*sin(60)*0.7},{2*0.75*0.7}) {};}
\foreach \k in {-4,-3,-1,0} {\node [shape=regular polygon,regular polygon sides=6,minimum size=0.7cm,fill=lightgray!50!white,inner sep=0,anchor=south,rotate=30] at ({(6-\k/2)*sin(60)*0.7},{\k*0.75*0.7}) {};}
\foreach \k in {1,2,4,5} {\node [shape=regular polygon,regular polygon sides=6,minimum size=0.7cm,fill=lightgray!50!white,inner sep=0,anchor=south,rotate=30] at ({(6+\k/2-0.5)*sin(60)*0.7},{2.25*0.7+\k*0.75*0.7}) {};}

% generating nodes

\foreach \j in {0,...,2}{
  \foreach \i in {0,...,8}{
  \node[hexa] (h\i;\j) at ({(\i-\j/2)*sin(60)*0.7},{\j*0.75*0.7}) {};}  }      
\foreach \j in {3,4}{ 
  \foreach \i in {0,...,8}{
  \node[hexa] (h\i;\j) at ({(\i+\j/2-2)*sin(60)*0.7},{\j*0.75*0.7}) {};}  } 
\foreach \j in {5,6,7,8}{ 
  \foreach \i in {4,...,8}{  
  \node[hexa] (h\i;\j) at ({(\i+\j/2-2)*sin(60)*0.7},{\j*0.75*0.7}) {};}  }
\foreach \j in {9}{ 
  \foreach \i in {4,...,7}{ 
  \node[hexa] (h\i;\j) at ({(\i+\j/2-2)*sin(60)*0.7},{\j*0.75*0.7}) {};}  }
\foreach \j in {10}{ 
  \foreach \i in {4,5}{  
  \node[hexa] (h\i;\j) at ({(\i+\j/2-2)*sin(60)*0.7},{\j*0.75*0.7}) {};}  }
  
\foreach \j in {0,1}{
  \node[hexa] (h-1;\j) at ({(-1-\j/2)*sin(60)*0.7},{\j*0.75*0.7}) {}; }
\foreach \j in {3,4}{
  \node[hexa] (h-1;\j) at ({(-1+(\j-3)/2-0.5)*sin(60)*0.7},{\j*0.75*0.7}) {}; } 
\foreach \j in {-4,...,-1}{
  \foreach \i in {4,...,8}{
  \node[hexa] (h\i;\j) at ({(\i-\j/2)*sin(60)*0.7},{\j*0.75*0.7}) {};}  } 
\foreach \j in {-5}{ 
  \foreach \i in {4,...,7}{
  \node[hexa] (h\i;\j) at ({(\i-\j/2)*sin(60)*0.7},{\j*0.75*0.7}) {};}  }   
\foreach \j in {-6}{ 
  \foreach \i in {4,5}{
  \node[hexa] (h\i;\j) at ({(\i-\j/2)*sin(60)*0.7},{\j*0.75*0.7}) {};}  }  

% Putting contents in the nodes

\foreach \k in {4,5}  {\node at (h\k;-6) {\begin{turn}{-60}$\cdots$\end{turn}};}

\foreach \k in {6,7}  {\node at (h\k;-5) {\begin{turn}{-60}$\cdots$\end{turn}};}
{\node at (h4;-5) {1};}
{\node [circle,draw,fill=gray,inner sep=3.5] at (h5;-5) {};}

{\node at (h8;-4) {\begin{turn}{-60}$\cdots$\end{turn}};}
{\node at (h4;-4) {1};}
{\node at (h5;-4) {3};}
{\node at (h6;-4) {$\boldsymbol{x}$};}
{\node [circle,draw,fill=gray,inner sep=3.5] at (h7;-4) {};}

{\node at (h8;-3) {1};}
{\node at (h4;-3) {2};}
{\node at (h7;-3) {3};}
{\node at (h6;-3) {$\boldsymbol{x'}$};}
{\node [circle,draw,fill=gray,inner sep=3.5] at (h5;-3) {};}

\foreach \k in {4,8}  {\node at (h\k;-2) {1};}
{\node at (h6;-2) {5};}
\foreach \k in {5,7} {\node [circle,draw,fill=gray,inner sep=3.5] at (h\k;-2) {};}

{\node at (h4;-1) {1};}
{\node at (h8;-1) {2};}
{\node at (h5;-1) {3};}
{\node at (h6;-1) {$\boldsymbol{x}$};}
{\node [circle,draw,fill=gray,inner sep=3.5] at (h7;-1) {};}

\foreach \j in {0,1,3,4}  {\node at (h-1;\j) {$\cdots$};}
\foreach \k in {1,2} {
  \foreach \j in {0,4}
  {\node at (h\k;\j) {1};} }
\foreach \k in {0,3,4} {
  \foreach \j in {0,4}
  {\node at (h\k;\j) {2};} }
  
{\node at (h8;0) {1};}
{\node at (h7;0) {3};}
{\node at (h6;0) {$\boldsymbol{x'}$};}
{\node [circle,draw,fill=gray,inner sep=3.5] at (h5;0) {};}

{\node at (h8;4) {1};}
{\node at (h7;4) {3};}
{\node at (h6;4) {$\boldsymbol{x'}$};}
{\node [circle,draw,fill=gray,inner sep=3.5] at (h5;4) {};}

\foreach \k in {0,1,3,4,7} {
  \foreach \j in {1,3}
  {\node [circle,draw,fill=gray,inner sep=3.5] at (h\k;\j) {};} }
\foreach \k in {2,5} {
  \foreach \j in {1,3}
  {\node at (h\k;\j) {3};} }
\foreach \j in {1,3}  {\node at (h6;\j) {4};}
\foreach \j in {1,3}  {\node at (h8;\j) {1};}

\foreach \k in {0}  {\node at (h\k;2) {$\cdots$};}
\foreach \k in {1,4}  {\node at (h\k;2) {5};}
\foreach \k in {2,5}  {\node at (h\k;2) {$\boldsymbol{x'}$};}
\foreach \k in {3,6}  {\node at (h\k;2) {$\boldsymbol{x}$};}
\foreach \k in {7} {\node [circle,draw,fill=gray,inner sep=3.5] at (h\k;2) {};}
\foreach \k in {8}  {\node at (h\k;2) {3};}

\foreach \k in {4,5}  {\node at (h\k;10) {\begin{turn}{60}$\cdots$\end{turn}};}

\foreach \k in {6,7}  {\node at (h\k;9) {\begin{turn}{60}$\cdots$\end{turn}};}
\foreach \k in {4}  {\node at (h\k;9) {1};}
\foreach \k in {5} {\node [circle,draw,fill=gray,inner sep=3.5] at (h\k;9) {};}

\foreach \k in {8}  {\node at (h\k;8) {\begin{turn}{60}$\cdots$\end{turn}};}
\foreach \k in {4}  {\node at (h\k;8) {1};}
\foreach \k in {5}  {\node at (h\k;8) {3};}
\foreach \k in {6}  {\node at (h\k;8) {$\boldsymbol{x}$};}
\foreach \k in {7} {\node [circle,draw,fill=gray,inner sep=3.5] at (h\k;8) {};}

\foreach \k in {8}  {\node at (h\k;7) {1};}
\foreach \k in {4}  {\node at (h\k;7) {2};}
\foreach \k in {7}  {\node at (h\k;7) {3};}
\foreach \k in {6}  {\node at (h\k;7) {$\boldsymbol{x'}$};}
\foreach \k in {5} {\node [circle,draw,fill=gray,inner sep=3.5] at (h\k;7) {};}

\foreach \k in {4,8}  {\node at (h\k;6) {1};}
\foreach \k in {6}  {\node at (h\k;6) {5};}
\foreach \k in {5,7} {\node [circle,draw,fill=gray,inner sep=3.5] at (h\k;6) {};}

\foreach \k in {4}  {\node at (h\k;5) {1};}
\foreach \k in {8}  {\node at (h\k;5) {2};}
\foreach \k in {5}  {\node at (h\k;5) {3};}
\foreach \k in {6}  {\node at (h\k;5) {$\boldsymbol{x}$};}
\foreach \k in {7} {\node [circle,draw,fill=gray,inner sep=3.5] at (h\k;5) {};}

% arrows

{\node at ({((-0.3)-(-0.8)/2)*sin(60)*0.7},{-0.8*0.75*0.7}) {$\boldsymbol{x}$};}
\draw [->] ({((0.2)-(-0.8)/2)*sin(60)*0.7},{-0.8*0.75*0.7}) to [bend left] ({((0.2)-(-0.8)/2)*sin(60)*0.7+2},{-0.8*0.75*0.7-1.1});

{\node at ({((-0.3)-(-0.8)/2)*sin(60)*0.7},{5.8*0.75*0.7}) {$\boldsymbol{x}$};}
\draw [->] ({((0.2)-(-0.8)/2)*sin(60)*0.7},{5.8*0.75*0.7}) to [bend right] ({((0.2)-(-0.8)/2)*sin(60)*0.7+2},{5.8*0.75*0.7+1.1});

\end{tikzpicture}
\caption{A splitter by merging two curves}\label{splitter}
\end{figure}
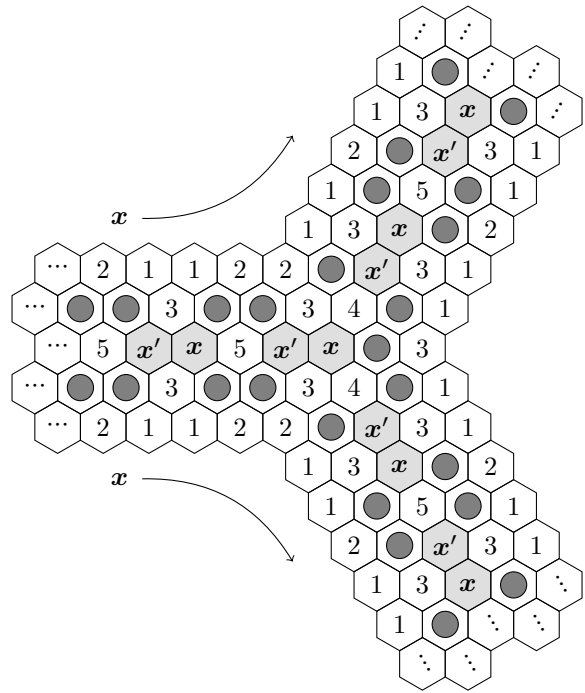

\newpage

%Figure~\ref{wire} represents the phase, therefore it is the improved form of Bondt's wire. 
The figure below represents the NOT gate.

%%%%%%%%%%%%%%%%%%%%%%%%%%
% Figure 11 : A NOT gate %
%%%%%%%%%%%%%%%%%%%%%%%%%%

\begin{figure}[ht]
\centering
$\boldsymbol{x} \xrightarrow{\hspace*{1.5cm}}\qquad\qquad\qquad\qquad\qquad \boldsymbol{x'} \xrightarrow{\hspace*{1.5cm}}$
\begin{tikzpicture} [hexa/.style= {shape=regular polygon,regular polygon sides=6,minimum size=0.7cm, draw,inner sep=0,anchor=south,rotate=30}]

% fill in some colors

\foreach \k in {2,3,5,6,8,10,11,13,14} {\node [shape=regular polygon,regular polygon sides=6,minimum size=0.7cm,fill=lightgray!50!white,inner sep=0,anchor=south,rotate=30] at ({(\k-2/2)*sin(60)*0.7},{2*0.75*0.7}) {};} 

% generating nodes

\foreach \j in {0,...,2}{ 
  \foreach \i in {0,...,15}{
  \node[hexa] (h\i;\j) at ({(\i-\j/2)*sin(60)*0.7},{\j*0.75*0.7}) {};}  }      
\foreach \j in {3,4}{
  \foreach \i in {0,...,15}{ 
  \node[hexa] (h\i;\j) at ({(\i+\j/2-2)*sin(60)*0.7},{\j*0.75*0.7}) {};}  } 
  
\node[hexa] (h{-1};0) at ({(-1)*sin(60)*0.7},{0*0.75*0.7}) {};
\node[hexa] (h{-1};4) at ({(-1+1/2-0.5)*sin(60)*0.7},{2.25*0.7+1*0.75*0.7}) {};
\node[hexa] (h{16};2) at ({(16-2/2)*sin(60)*0.7},{2*0.75*0.7}) {};

% putting in contents

\foreach \k in {2,5,8,11,14}  {\node at (h\k;2) {$\boldsymbol{x'}$};}
\foreach \k in {3,6,10,13}  {\node at (h\k;2) {$\boldsymbol{x}$};}
\foreach \k in {1,4,7,9,12,15}  {\node at (h\k;2) {5};}
\foreach \k in {0,3,6,7,8,11,14} {\node at (h\k;0) {2};}
\foreach \k in {1,2,4,5,9,10,12,13} {\node at (h\k;0) {1};}
\foreach \k in {1,2,4,5,9,10,12,13} {\node at (h\k;4) {1};}
\foreach \k in {0,3,6,7,8,11,14} {\node at (h\k;4) {2};}
\foreach \k in {2,5,10,13} {\node at (h\k;1) {3};}
\foreach \k in {2,5,10,13} {\node at (h\k;3) {3};}
\foreach \k in {1,3,4,6,7,8,9,11,12,14} {\node [circle,draw,fill=gray,inner sep=3.5] at (h\k;1) {};}
\foreach \k in {1,3,4,6,7,8,9,11,12,14} {\node [circle,draw,fill=gray,inner sep=3.5] at (h\k;3) {};}
\foreach \k in {1,2,3}  {\node at (h0;\k) {$\cdots$};}
\foreach \k in {0,1,3,4}  {\node at (h15;\k) {$\cdots$};}
  {\node at (h{16};2) {$\cdots$};}
  {\node at (h{-1};0) {$\cdots$};}
  {\node at (h{-1};4) {$\cdots$};}
\end{tikzpicture}
\caption{A NOT gate on a hexagonal Minesweeper}\label{notgate}
\end{figure}
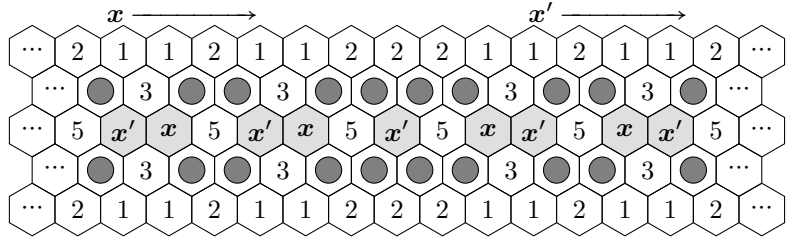

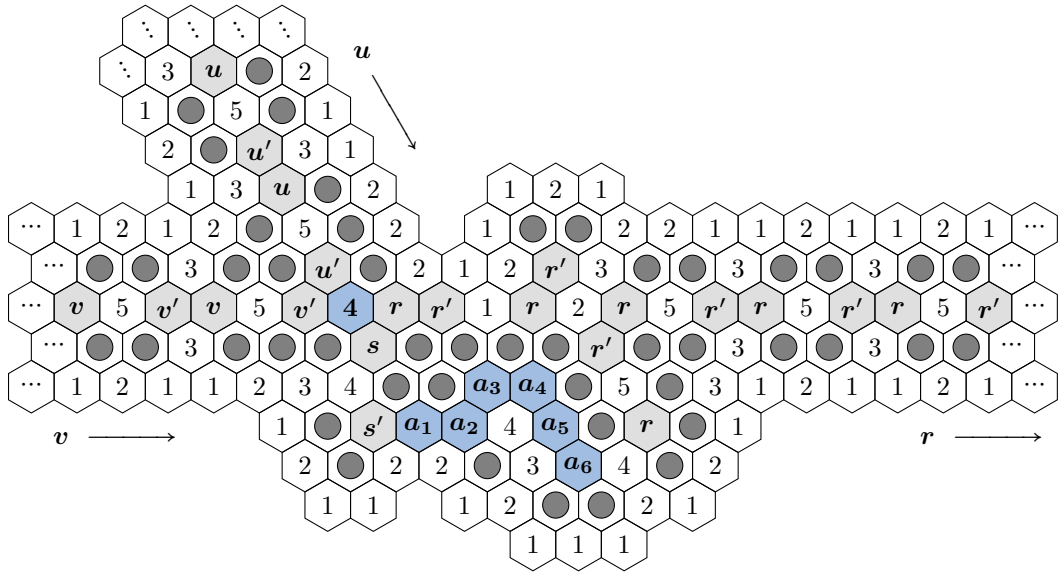
\begin{figure}[ht]
\centering
\begin{tikzpicture} [hexa/.style= {shape=regular polygon,regular polygon sides=6,minimum size=0.7cm, draw,inner sep=0,anchor=south,rotate=30}]

% Fill in some color

\foreach \k in {10} {\node [shape=regular polygon,regular polygon sides=6,minimum size=0.7cm,fill=RoyalBlue!35!white,inner sep=0,anchor=south,rotate=30] at ({(\k-(-2)/2)*sin(60)*0.7},{-2*0.75*0.7}) {};} 
\foreach \k in {7,8,10} {\node [shape=regular polygon,regular polygon sides=6,minimum size=0.7cm,fill=RoyalBlue!35!white,inner sep=0,anchor=south,rotate=30] at ({(\k-(-1)/2)*sin(60)*0.7},{-1*0.75*0.7}) {};} 
\foreach \k in {6,12} {\node [shape=regular polygon,regular polygon sides=6,minimum size=0.7cm,fill=lightgray!50!white,inner sep=0,anchor=south,rotate=30] at ({(\k-(-1)/2)*sin(60)*0.7},{-1*0.75*0.7}) {};}
\foreach \k in {9,10} {\node [shape=regular polygon,regular polygon sides=6,minimum size=0.7cm,fill=RoyalBlue!35!white,inner sep=0,anchor=south,rotate=30] at ({(\k-(0)/2)*sin(60)*0.7},{0*0.75*0.7}) {};}
\foreach \k in {7,12} {\node [shape=regular polygon,regular polygon sides=6,minimum size=0.7cm,fill=lightgray!50!white,inner sep=0,anchor=south,rotate=30] at ({(\k-(1)/2)*sin(60)*0.7},{1*0.75*0.7}) {};}
\foreach \k in {1,3,4,6,8,9,11,13,15,16,18,19,21} {\node [shape=regular polygon,regular polygon sides=6,minimum size=0.7cm,fill=lightgray!50!white,inner sep=0,anchor=south,rotate=30] at ({(\k-(2)/2)*sin(60)*0.7},{2*0.75*0.7}) {};}
\foreach \k in {7} {\node [shape=regular polygon,regular polygon sides=6,minimum size=0.7cm,fill=RoyalBlue!35!white,inner sep=0,anchor=south,rotate=30] at ({(\k-(2)/2)*sin(60)*0.7},{2*0.75*0.7}) {};}
\foreach \k in {6,11} {\node [shape=regular polygon,regular polygon sides=6,minimum size=0.7cm,fill=lightgray!50!white,inner sep=0,anchor=south,rotate=30] at ({((\k+1)-(3)/2)*sin(60)*0.7},{3*0.75*0.7}) {};}
\foreach \k in {4} {\node [shape=regular polygon,regular polygon sides=6,minimum size=0.7cm,fill=lightgray!50!white,inner sep=0,anchor=south,rotate=30] at ({((\k+3)-(5)/2)*sin(60)*0.7},{5*0.75*0.7}) {};}
\foreach \k in {3} {\node [shape=regular polygon,regular polygon sides=6,minimum size=0.7cm,fill=lightgray!50!white,inner sep=0,anchor=south,rotate=30] at ({((\k+4)-(6)/2)*sin(60)*0.7},{6*0.75*0.7}) {};}
\foreach \k in {1} {\node [shape=regular polygon,regular polygon sides=6,minimum size=0.7cm,fill=lightgray!50!white,inner sep=0,anchor=south,rotate=30] at ({((\k+6)-(8)/2)*sin(60)*0.7},{8*0.75*0.7}) {};}

% Generating nodes

\foreach \j in {0,...,2}{
  \foreach \i in {0,...,21}{
  \node[hexa] (h\i;\j) at ({(\i-\j/2)*sin(60)*0.7},{\j*0.75*0.7}) {};}  }  

\foreach \i in {4,...,14}{  \node[hexa] (h\i;-1) at ({(\i+1/2)*sin(60)*0.7},{-1*0.75*0.7}) {};}
\foreach \i in {4,...,13}{  \node[hexa] (h\i;-2) at ({(\i+2/2)*sin(60)*0.7},{-2*0.75*0.7}) {};}
\foreach \i in {4,5,7,8,9,10,11,12}{  \node[hexa] (h\i;-3) at ({(\i+3/2)*sin(60)*0.7},{-3*0.75*0.7}) {};}
\foreach \i in {8,9,10}{  \node[hexa] (h\i;-4) at ({(\i+4/2)*sin(60)*0.7},{-4*0.75*0.7}) {};}
        
\foreach \j in {3}{
  \foreach \i in {0,...,21}{  
  \node[hexa] (h\i;\j) at ({(\i+\j/2-2)*sin(60)*0.7},{\j*0.75*0.7}) {};}  } 
  
\foreach \j in {4}{ 
  \foreach \i in {0,...,7}{  
  \node[hexa] (h\i;\j) at ({(\i+\j/2-2)*sin(60)*0.7},{\j*0.75*0.7}) {};}  } 

\foreach \j in {4}{
  \foreach \i in {9,...,21}{  
  \node[hexa] (h\i;\j) at ({(\i+\j/2-2)*sin(60)*0.7},{\j*0.75*0.7}) {};}  } 
   
\node[hexa] (h-1;0) at ({(-1)*sin(60)*0.7},{0*0.75*0.7}) {};
\node[hexa] (h-1;4) at ({(-1+1/2-0.5)*sin(60)*0.7},{2.25*0.7+1*0.75*0.7}) {};
\node[hexa] (h22;2) at ({(22-2/2)*sin(60)*0.7},{2*0.75*0.7}) {};

\foreach \i in {2,3,4,5,6,9,10,11}{
  \node[hexa] (h\i;5) at ({(\i+2/2-0.5)*sin(60)*0.7},{2.25*0.7+2*0.75*0.7}) {};}
\foreach \i in {1,...,5}{
  \node[hexa] (h\i;6) at ({(\i+3/2-0.5)*sin(60)*0.7},{2.25*0.7+3*0.75*0.7}) {};}
\foreach \i in {0,...,4}{
  \node[hexa] (h\i;7) at ({(\i+4/2-0.5)*sin(60)*0.7},{2.25*0.7+4*0.75*0.7}) {};}
\foreach \i in {-1,...,3}{
  \node[hexa] (h\i;8) at ({(\i+5/2-0.5)*sin(60)*0.7},{2.25*0.7+5*0.75*0.7}) {};}
\foreach \i in {-1,...,2}{
  \node[hexa] (h\i;9) at ({(\i+6/2-0.5)*sin(60)*0.7},{2.25*0.7+6*0.75*0.7}) {};}
 
% Putting contents in the nodes

\foreach \k in {1,2,3}  {\node at (h0;\k) {$\cdots$};}
\foreach \k in {0,1,3,4}  {\node at (h21;\k) {$\cdots$};}
  {\node at (h22;2) {$\cdots$};}
  {\node at (h-1;0) {$\cdots$};}
  {\node at (h-1;4) {$\cdots$};}
  {\node at (h-1;8) {\begin{turn}{-60}$\cdots$\end{turn}};}
\foreach \k in {-1,...,2}  {\node at (h\k;9) {\begin{turn}{-60}$\cdots$\end{turn}};}

\foreach \k in {8,9,10}  {\node at (h\k;-4) {1};}

\foreach \k in {4,5,7,12}  {\node at (h\k;-3) {1};}
\foreach \k in {8,11}  {\node at (h\k;-3) {2};}
\foreach \k in {9,10}  {\node [circle,draw,fill=gray,inner sep=3.5] at (h\k;-3) {};}

\foreach \k in {4,6,7,13}  {\node at (h\k;-2) {2};}
\foreach \k in {9}  {\node at (h\k;-2) {3};}
\foreach \k in {11}  {\node at (h\k;-2) {4};}
\foreach \k in {10}  {\node at (h\k;-2) {$\boldsymbol{a_6}$};}
\foreach \k in {5,8,12}  {\node [circle,draw,fill=gray,inner sep=3.5] at (h\k;-2) {};}

\foreach \k in {4,14}  {\node at (h\k;-1) {1};}
{\node at (h6;-1) {$\boldsymbol{s'}$};}
{\node at (h7;-1) {$\boldsymbol{a_1}$};}
{\node at (h8;-1) {$\boldsymbol{a_2}$};}
{\node at (h9;-1) {4};}
{\node at (h10;-1) {$\boldsymbol{a_5}$};}
{\node at (h12;-1) {$\boldsymbol{r}$};}
\foreach \k in {5,11,13}  {\node [circle,draw,fill=gray,inner sep=3.5] at (h\k;-1) {};}

\foreach \k in {0,2,3,15,17,18,20}  {\node at (h\k;0) {1};}
\foreach \k in {1,4,16,19}  {\node at (h\k;0) {2};}
\foreach \k in {5,14}  {\node at (h\k;0) {3};}
\foreach \k in {6}  {\node at (h\k;0) {4};}
\foreach \k in {12}  {\node at (h\k;0) {5};}
\foreach \k in {9}  {\node at (h\k;0) {$\boldsymbol{a_3}$};}
\foreach \k in {10}  {\node at (h\k;0) {$\boldsymbol{a_4}$};} 
\foreach \k in {7,8,11,13}  {\node [circle,draw,fill=gray,inner sep=3.5] at (h\k;0) {};} 

\foreach \k in {3,15,18}  {\node at (h\k;1) {3};}
\foreach \k in {1,2,4,5,6,8,9,10,11,13,14,16,17,19,20}  {\node [circle,draw,fill=gray,inner sep=3.5] at (h\k;1) {};}
\foreach \k in {7}  {\node at (h\k;1) {$\boldsymbol{s}$};}
\foreach \k in {12}  {\node at (h\k;1) {$\boldsymbol{r'}$};}

\foreach \k in {10}  {\node at (h\k;2) {1};}
\foreach \k in {12}  {\node at (h\k;2) {2};}
\foreach \k in {7}  {\node at (h\k;2) {{\textbf{4}}};}
\foreach \k in {2,5,14,17,20}  {\node at (h\k;2) {5};}
\foreach \k in {1,4}  {\node at (h\k;2) {$\boldsymbol{v}$};}
\foreach \k in {3,6}  {\node at (h\k;2) {$\boldsymbol{v'}$};}
\foreach \k in {8,11,13,16,19}  {\node at (h\k;2) {$\boldsymbol{r}$};}
\foreach \k in {9,15,18,21}  {\node at (h\k;2) {$\boldsymbol{r'}$};}

\foreach \k in {9}  {\node at (h\k;3) {1};}
\foreach \k in {8,10}  {\node at (h\k;3) {2};}
\foreach \k in {3,12,15,18}  {\node at (h\k;3) {3};}
\foreach \k in {1,2,4,5,7,13,14,16,17,19,20}  {\node [circle,draw,fill=gray,inner sep=3.5] at (h\k;3) {};} 
\foreach \k in {11}  {\node at (h\k;3) {$\boldsymbol{r'}$};}
\foreach \k in {6}  {\node at (h\k;3) {$\boldsymbol{u'}$};}

\foreach \k in {0,2,9,14,15,17,18,20}  {\node at (h\k;4) {1};}
\foreach \k in {1,3,7,12,13,16,19}  {\node at (h\k;4) {2};}
\foreach \k in {5}  {\node at (h\k;4) {5};}
\foreach \k in {4,6,10,11}  {\node [circle,draw,fill=gray,inner sep=3.5] at (h\k;4) {};}

\foreach \k in {2,9,11}  {\node at (h\k;5) {1};}
\foreach \k in {6,10}  {\node at (h\k;5) {2};}
\foreach \k in {3}  {\node at (h\k;5) {3};}
\foreach \k in {4}  {\node at (h\k;5) {$\boldsymbol{u}$};}
\foreach \k in {5}  {\node [circle,draw,fill=gray,inner sep=3.5] at (h\k;5) {};}

\foreach \k in {5}  {\node at (h\k;6) {1};}
\foreach \k in {1}  {\node at (h\k;6) {2};}
\foreach \k in {4}  {\node at (h\k;6) {3};}
\foreach \k in {3}  {\node at (h\k;6) {$\boldsymbol{u'}$};}
\foreach \k in {2}  {\node [circle,draw,fill=gray,inner sep=3.5] at (h\k;6) {};}

\foreach \k in {0,4}  {\node at (h\k;7) {1};}
\foreach \k in {2}  {\node at (h\k;7) {5};}
\foreach \k in {1,3}  {\node [circle,draw,fill=gray,inner sep=3.5] at (h\k;7) {};}

\foreach \k in {3}  {\node at (h\k;8) {2};}
\foreach \k in {0}  {\node at (h\k;8) {3};}
\foreach \k in {1}  {\node at (h\k;8) {$\boldsymbol{u}$};}
\foreach \k in {2}  {\node [circle,draw,fill=gray,inner sep=3.5] at (h\k;8) {};}

% arrows

{\node at ({((10.5)-(9)/2)*sin(60)*0.7},{9*0.75*0.7}) {$\boldsymbol{u}$};}
{\node at ({((10.5)-(7.5)/2)*sin(60)*0.7},{7.5*0.75*0.7}) {\begin{turn}{-60}$\xrightarrow{\hspace*{1cm}}$\end{turn}};}

{\node at ({((-1)-(-0.8)/2)*sin(60)*0.7},{-0.8*0.75*0.7}) {$\boldsymbol{v}$};}
{\node at ({((0.6)-(-0.7)/2)*sin(60)*0.7},{-0.7*0.75*0.7}) {$\xrightarrow{\hspace*{1cm}}$};}

{\node at ({((18)-(-0.8)/2)*sin(60)*0.7},{-0.8*0.75*0.7}) {$\boldsymbol{r}$};}
{\node at ({((19.6)-(-0.7)/2)*sin(60)*0.7},{-0.7*0.75*0.7}) {$\xrightarrow{\hspace*{1cm}}$};}
  
\end{tikzpicture}
\caption{An OR gate on a hexagonal Minesweeper}\label{orgate}
\end{figure}

Duplicating two NOT gates, we can make a phase-changer easily (in this paper, I do not provide such figure as it is logically simple to determine).
Now, I give an OR gate and an AND gate on a hexagonal grid. Figure~\ref{orgate} is an OR gate.

%%%%%%%%%%%%%%%%%%%%%%%
% Figure 12 : OR Gate %
%%%%%%%%%%%%%%%%%%%%%%%

%%%%%%%%%%%%%%%%%%%% End of OR Gate %%%%%%%%%%%%%%%%%%%%%%%%

As the figure seems quite complicated, it requires some explanation for it being an OR gate. Let us prove this case-by-case.

\begin{case}
(both $\boldsymbol{u}$ and $\boldsymbol{v}$ are T)\\
Since $\boldsymbol{u'}$ and $\boldsymbol{v'}$ are both F, $\boldsymbol{r}$ and $\boldsymbol{s}$ should be both T. We can easily check that $\boldsymbol{a_1}$, $\boldsymbol{a_3}$, $\boldsymbol{a_4}$, and $\boldsymbol{a_5}$ are T and the rest $\boldsymbol{a_i}$'s are F.
\end{case}

\begin{case}
(only one of $\boldsymbol{u}$ and $\boldsymbol{v}$ is T)\\
Investigating the blue `4', only one between $\boldsymbol{r}$ and $\boldsymbol{s}$ is T. If $\boldsymbol{r}$ is F and $\boldsymbol{s}$ is T, then $\boldsymbol{a_1}$ is T and $\boldsymbol{a_2}$ is F. Now we are looking through $\boldsymbol{a_1}$ to $\boldsymbol{a_6}$. Since $\boldsymbol{a_2}$ is F, $\boldsymbol{a_3}$, $\boldsymbol{a_4}$, $\boldsymbol{a_5}$ are all T. Therefore $\boldsymbol{a_6}$ should be F. But then $\boldsymbol{r}$ should be T, which is a contradiction! Therefore $\boldsymbol{r}$ should be T and $\boldsymbol{s}$ is F. Then $\boldsymbol{a_1}$ and $\boldsymbol{a_6}$ are F and $\boldsymbol{a_2}$, $\boldsymbol{a_5}$ are T. Since both $\boldsymbol{a_2}$, $\boldsymbol{a_5}$ are T, only one between $\boldsymbol{a_3}$, $\boldsymbol{a_4}$ is T(we cannot decide which one should be T).
\end{case}

\begin{case}
(both $\boldsymbol{u}$ and $\boldsymbol{v}$ are F)\\
Since $\boldsymbol{u'}$ and $\boldsymbol{v'}$ are both T, $\boldsymbol{r}$ and $\boldsymbol{s}$ should be both F. We can easily check that $\boldsymbol{a_2}$, $\boldsymbol{a_3}$, $\boldsymbol{a_4}$, and $\boldsymbol{a_6}$ are T and the rest $\boldsymbol{a_i}$'s are F.
\end{case}

By examining the {\textit{Case}} 1 to 3, we can finally conclude that figure~\ref{orgate} is actually an OR gate. The truth table for this OR gate is as below:

\begin{table}[!htbp]
\centering
\caption{Truth table for the OR gate on a hexagonal Minesweeper}
\vskip 0.3em
\begin{tabular}{|c|c||c|c|c|c|c|c|c||c|}
\hline
$\boldsymbol{u}$ & $\boldsymbol{v}$ & $\boldsymbol{s}$ & $\boldsymbol{a_1}$ & $\boldsymbol{a_2}$ & $\boldsymbol{a_3}$ & $\boldsymbol{a_4}$ & $\boldsymbol{a_5}$ & $\boldsymbol{a_6}$ & $\boldsymbol{r}=\boldsymbol{u}\vee\boldsymbol{v}$\\\hline\hline
T&T&T&T&F&T&T&T&F&T\\\hline
T&F&F&F&T&\multicolumn{2}{c|}{TF or FT}&T&F&T\\\hline
F&T&F&F&T&\multicolumn{2}{c|}{TF or FT}&T&F&T\\\hline
F&F&F&F&T&T&T&F&T&F\\\hline
\end{tabular}
\label{tab:ortable}
\end{table}

By changing the position of $\boldsymbol{a_1}$ through $\boldsymbol{a_6}$, we can make an AND gate similarly.

%%%%%%%%%%%%%%%%%%%%%%%%
% Figure 13 : AND Gate %
%%%%%%%%%%%%%%%%%%%%%%%%

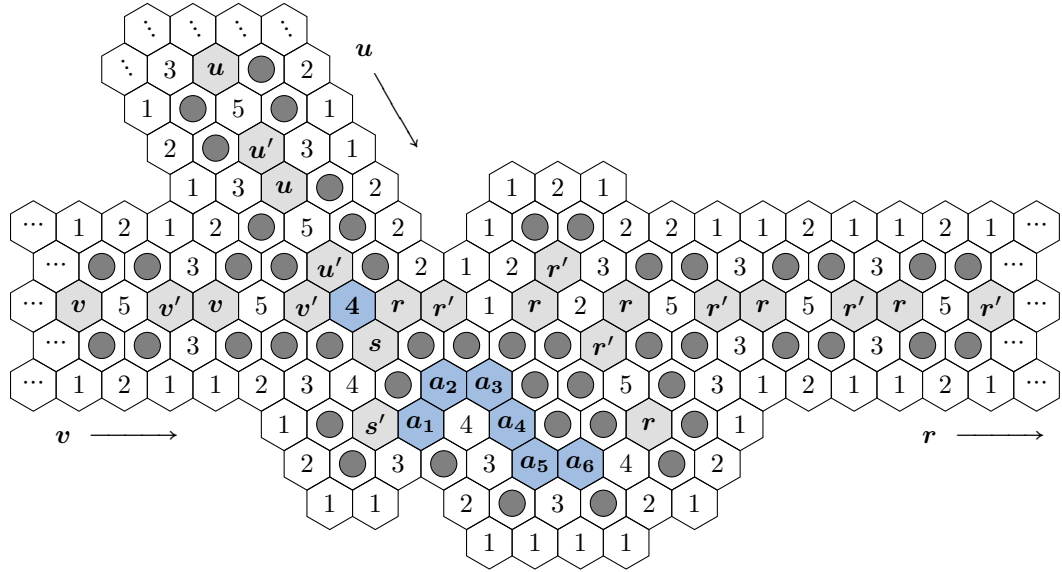
\begin{figure}[ht]
\centering
\begin{tikzpicture} [hexa/.style= {shape=regular polygon,regular polygon sides=6,minimum size=0.7cm, draw,inner sep=0,anchor=south,rotate=30}]

% Fill in some color

\foreach \k in {9,10} {\node [shape=regular polygon,regular polygon sides=6,minimum size=0.7cm,fill=RoyalBlue!35!white,inner sep=0,anchor=south,rotate=30] at ({(\k-(-2)/2)*sin(60)*0.7},{-2*0.75*0.7}) {};} 
\foreach \k in {7,9} {\node [shape=regular polygon,regular polygon sides=6,minimum size=0.7cm,fill=RoyalBlue!35!white,inner sep=0,anchor=south,rotate=30] at ({(\k-(-1)/2)*sin(60)*0.7},{-1*0.75*0.7}) {};} 
\foreach \k in {6,12} {\node [shape=regular polygon,regular polygon sides=6,minimum size=0.7cm,fill=lightgray!50!white,inner sep=0,anchor=south,rotate=30] at ({(\k-(-1)/2)*sin(60)*0.7},{-1*0.75*0.7}) {};}
\foreach \k in {8,9} {\node [shape=regular polygon,regular polygon sides=6,minimum size=0.7cm,fill=RoyalBlue!35!white,inner sep=0,anchor=south,rotate=30] at ({(\k-(0)/2)*sin(60)*0.7},{0*0.75*0.7}) {};}
\foreach \k in {7,12} {\node [shape=regular polygon,regular polygon sides=6,minimum size=0.7cm,fill=lightgray!50!white,inner sep=0,anchor=south,rotate=30] at ({(\k-(1)/2)*sin(60)*0.7},{1*0.75*0.7}) {};}
\foreach \k in {1,3,4,6,8,9,11,13,15,16,18,19,21} {\node [shape=regular polygon,regular polygon sides=6,minimum size=0.7cm,fill=lightgray!50!white,inner sep=0,anchor=south,rotate=30] at ({(\k-(2)/2)*sin(60)*0.7},{2*0.75*0.7}) {};}
\foreach \k in {7} {\node [shape=regular polygon,regular polygon sides=6,minimum size=0.7cm,fill=RoyalBlue!35!white,inner sep=0,anchor=south,rotate=30] at ({(\k-(2)/2)*sin(60)*0.7},{2*0.75*0.7}) {};}
\foreach \k in {6,11} {\node [shape=regular polygon,regular polygon sides=6,minimum size=0.7cm,fill=lightgray!50!white,inner sep=0,anchor=south,rotate=30] at ({((\k+1)-(3)/2)*sin(60)*0.7},{3*0.75*0.7}) {};}
\foreach \k in {4} {\node [shape=regular polygon,regular polygon sides=6,minimum size=0.7cm,fill=lightgray!50!white,inner sep=0,anchor=south,rotate=30] at ({((\k+3)-(5)/2)*sin(60)*0.7},{5*0.75*0.7}) {};}
\foreach \k in {3} {\node [shape=regular polygon,regular polygon sides=6,minimum size=0.7cm,fill=lightgray!50!white,inner sep=0,anchor=south,rotate=30] at ({((\k+4)-(6)/2)*sin(60)*0.7},{6*0.75*0.7}) {};}
\foreach \k in {1} {\node [shape=regular polygon,regular polygon sides=6,minimum size=0.7cm,fill=lightgray!50!white,inner sep=0,anchor=south,rotate=30] at ({((\k+6)-(8)/2)*sin(60)*0.7},{8*0.75*0.7}) {};}

% Generating nodes

\foreach \j in {0,...,2}{
  \foreach \i in {0,...,21}{
  \node[hexa] (h\i;\j) at ({(\i-\j/2)*sin(60)*0.7},{\j*0.75*0.7}) {};}  }  

\foreach \i in {4,...,14}{
  \node[hexa] (h\i;-1) at ({(\i+1/2)*sin(60)*0.7},{-1*0.75*0.7}) {};}
\foreach \i in {4,...,13}{
  \node[hexa] (h\i;-2) at ({(\i+2/2)*sin(60)*0.7},{-2*0.75*0.7}) {};}
\foreach \i in {4,5,7,8,9,10,11,12}{
  \node[hexa] (h\i;-3) at ({(\i+3/2)*sin(60)*0.7},{-3*0.75*0.7}) {};}
\foreach \i in {7,8,9,10}{
  \node[hexa] (h\i;-4) at ({(\i+4/2)*sin(60)*0.7},{-4*0.75*0.7}) {};}
          
\foreach \j in {3}{
  \foreach \i in {0,...,21}{ 
  \node[hexa] (h\i;\j) at ({(\i+\j/2-2)*sin(60)*0.7},{\j*0.75*0.7}) {};}  } 
  
\foreach \j in {4}{ 
  \foreach \i in {0,...,7}{  
  \node[hexa] (h\i;\j) at ({(\i+\j/2-2)*sin(60)*0.7},{\j*0.75*0.7}) {};}  } 

\foreach \j in {4}{
  \foreach \i in {9,...,21}{ 
  \node[hexa] (h\i;\j) at ({(\i+\j/2-2)*sin(60)*0.7},{\j*0.75*0.7}) {};}  } 

\node[hexa] (h-1;0) at ({(-1)*sin(60)*0.7},{0*0.75*0.7}) {};
\node[hexa] (h-1;4) at ({(-1+1/2-0.5)*sin(60)*0.7},{2.25*0.7+1*0.75*0.7}) {};
\node[hexa] (h22;2) at ({(22-2/2)*sin(60)*0.7},{2*0.75*0.7}) {};

\foreach \i in {2,3,4,5,6,9,10,11}{
  \node[hexa] (h\i;5) at ({(\i+2/2-0.5)*sin(60)*0.7},{2.25*0.7+2*0.75*0.7}) {};}
\foreach \i in {1,...,5}{
  \node[hexa] (h\i;6) at ({(\i+3/2-0.5)*sin(60)*0.7},{2.25*0.7+3*0.75*0.7}) {};}
\foreach \i in {0,...,4}{
  \node[hexa] (h\i;7) at ({(\i+4/2-0.5)*sin(60)*0.7},{2.25*0.7+4*0.75*0.7}) {};}
\foreach \i in {-1,...,3}{
  \node[hexa] (h\i;8) at ({(\i+5/2-0.5)*sin(60)*0.7},{2.25*0.7+5*0.75*0.7}) {};}
\foreach \i in {-1,...,2}{
  \node[hexa] (h\i;9) at ({(\i+6/2-0.5)*sin(60)*0.7},{2.25*0.7+6*0.75*0.7}) {};}

% Putting contents in the nodes

\foreach \k in {1,2,3}  {\node at (h0;\k) {$\cdots$};}
\foreach \k in {0,1,3,4}  {\node at (h21;\k) {$\cdots$};}
  {\node at (h22;2) {$\cdots$};}
  {\node at (h-1;0) {$\cdots$};}
  {\node at (h-1;4) {$\cdots$};}
  {\node at (h-1;8) {\begin{turn}{-60}$\cdots$\end{turn}};}
\foreach \k in {-1,...,2}  {\node at (h\k;9) {\begin{turn}{-60}$\cdots$\end{turn}};}

\foreach \k in {7,8,9,10}  {\node at (h\k;-4) {1};}

\foreach \k in {4,5,12}  {\node at (h\k;-3) {1};}
\foreach \k in {7,11}  {\node at (h\k;-3) {2};}
\foreach \k in {9}  {\node at (h\k;-3) {3};}
\foreach \k in {8,10}  {\node [circle,draw,fill=gray,inner sep=3.5] at (h\k;-3) {};}

\foreach \k in {4,13}  {\node at (h\k;-2) {2};}
\foreach \k in {6,8}  {\node at (h\k;-2) {3};}
\foreach \k in {11}  {\node at (h\k;-2) {4};}
\foreach \k in {9}  {\node at (h\k;-2) {$\boldsymbol{a_5}$};}
\foreach \k in {10}  {\node at (h\k;-2) {$\boldsymbol{a_6}$};}
\foreach \k in {5,7,12}  {\node [circle,draw,fill=gray,inner sep=3.5] at (h\k;-2) {};}

\foreach \k in {4,14}  {\node at (h\k;-1) {1};}
{\node at (h6;-1) {$\boldsymbol{s'}$};}
{\node at (h7;-1) {$\boldsymbol{a_1}$};}
{\node at (h8;-1) {4};}
{\node at (h9;-1) {$\boldsymbol{a_4}$};}
{\node at (h12;-1) {$\boldsymbol{r}$};}
\foreach \k in {5,10,11,13}  {\node [circle,draw,fill=gray,inner sep=3.5] at (h\k;-1) {};}

\foreach \k in {0,2,3,15,17,18,20}  {\node at (h\k;0) {1};}
\foreach \k in {1,4,16,19}  {\node at (h\k;0) {2};}
\foreach \k in {5,14}  {\node at (h\k;0) {3};}
\foreach \k in {6}  {\node at (h\k;0) {4};}
\foreach \k in {12}  {\node at (h\k;0) {5};}
\foreach \k in {8}  {\node at (h\k;0) {$\boldsymbol{a_2}$};}
\foreach \k in {9}  {\node at (h\k;0) {$\boldsymbol{a_3}$};} 
\foreach \k in {7,10,11,13}  {\node [circle,draw,fill=gray,inner sep=3.5] at (h\k;0) {};} 

\foreach \k in {3,15,18}  {\node at (h\k;1) {3};}
\foreach \k in {1,2,4,5,6,8,9,10,11,13,14,16,17,19,20}  {\node [circle,draw,fill=gray,inner sep=3.5] at (h\k;1) {};}
\foreach \k in {7}  {\node at (h\k;1) {$\boldsymbol{s}$};}
\foreach \k in {12}  {\node at (h\k;1) {$\boldsymbol{r'}$};}

\foreach \k in {10}  {\node at (h\k;2) {1};}
\foreach \k in {12}  {\node at (h\k;2) {2};}
\foreach \k in {7}  {\node at (h\k;2) {{\textbf{4}}};}
\foreach \k in {2,5,14,17,20}  {\node at (h\k;2) {5};}
\foreach \k in {1,4}  {\node at (h\k;2) {$\boldsymbol{v}$};}
\foreach \k in {3,6}  {\node at (h\k;2) {$\boldsymbol{v'}$};}
\foreach \k in {8,11,13,16,19}  {\node at (h\k;2) {$\boldsymbol{r}$};}
\foreach \k in {9,15,18,21}  {\node at (h\k;2) {$\boldsymbol{r'}$};}

\foreach \k in {9}  {\node at (h\k;3) {1};}
\foreach \k in {8,10}  {\node at (h\k;3) {2};}
\foreach \k in {3,12,15,18}  {\node at (h\k;3) {3};}
\foreach \k in {1,2,4,5,7,13,14,16,17,19,20}  {\node [circle,draw,fill=gray,inner sep=3.5] at (h\k;3) {};} 
\foreach \k in {11}  {\node at (h\k;3) {$\boldsymbol{r'}$};}
\foreach \k in {6}  {\node at (h\k;3) {$\boldsymbol{u'}$};}

\foreach \k in {0,2,9,14,15,17,18,20}  {\node at (h\k;4) {1};}
\foreach \k in {1,3,7,12,13,16,19}  {\node at (h\k;4) {2};}
\foreach \k in {5}  {\node at (h\k;4) {5};}
\foreach \k in {4,6,10,11}  {\node [circle,draw,fill=gray,inner sep=3.5] at (h\k;4) {};}

\foreach \k in {2,9,11}  {\node at (h\k;5) {1};}
\foreach \k in {6,10}  {\node at (h\k;5) {2};}
\foreach \k in {3}  {\node at (h\k;5) {3};}
\foreach \k in {4}  {\node at (h\k;5) {$\boldsymbol{u}$};}
\foreach \k in {5}  {\node [circle,draw,fill=gray,inner sep=3.5] at (h\k;5) {};}

\foreach \k in {5}  {\node at (h\k;6) {1};}
\foreach \k in {1}  {\node at (h\k;6) {2};}
\foreach \k in {4}  {\node at (h\k;6) {3};}
\foreach \k in {3}  {\node at (h\k;6) {$\boldsymbol{u'}$};}
\foreach \k in {2}  {\node [circle,draw,fill=gray,inner sep=3.5] at (h\k;6) {};}

\foreach \k in {0,4}  {\node at (h\k;7) {1};}
\foreach \k in {2}  {\node at (h\k;7) {5};}
\foreach \k in {1,3}  {\node [circle,draw,fill=gray,inner sep=3.5] at (h\k;7) {};}

\foreach \k in {3}  {\node at (h\k;8) {2};}
\foreach \k in {0}  {\node at (h\k;8) {3};}
\foreach \k in {1}  {\node at (h\k;8) {$\boldsymbol{u}$};}
\foreach \k in {2}  {\node [circle,draw,fill=gray,inner sep=3.5] at (h\k;8) {};}

% arrows

{\node at ({((10.5)-(9)/2)*sin(60)*0.7},{9*0.75*0.7}) {$\boldsymbol{u}$};}
{\node at ({((10.5)-(7.5)/2)*sin(60)*0.7},{7.5*0.75*0.7}) {\begin{turn}{-60}$\xrightarrow{\hspace*{1cm}}$\end{turn}};}

{\node at ({((-1)-(-0.8)/2)*sin(60)*0.7},{-0.8*0.75*0.7}) {$\boldsymbol{v}$};}
{\node at ({((0.6)-(-0.7)/2)*sin(60)*0.7},{-0.7*0.75*0.7}) {$\xrightarrow{\hspace*{1cm}}$};}

{\node at ({((18)-(-0.8)/2)*sin(60)*0.7},{-0.8*0.75*0.7}) {$\boldsymbol{r}$};}
{\node at ({((19.6)-(-0.7)/2)*sin(60)*0.7},{-0.7*0.75*0.7}) {$\xrightarrow{\hspace*{1cm}}$};}
  
\end{tikzpicture}
\caption{An AND gate on a hexagonal Minesweeper}\label{andgate}
\end{figure}

%%%%%%%%%%%%%%%%%%%% End of AND Gate %%%%%%%%%%%%%%%%%%%%%%%%

The reason that figure~\ref{andgate} is actually an AND gate is very similar to that of the OR gate shown earlier in this paper. Notice that by examining through $\boldsymbol{a_1}$ to $\boldsymbol{a_6}$, we can easily check that the result $\boldsymbol{r}$ is only T when two inputs $\boldsymbol{u}$ and $\boldsymbol{v}$ are both T. The truth table for this AND gate is as below:

\begin{table}[!htbp]
\centering
\caption{Truth table for the AND gate on a hexagonal Minesweeper}
\vskip 0.3em
\begin{tabular}{|c|c||c|c|c|c|c|c|c||c|}
\hline
$\boldsymbol{u}$ & $\boldsymbol{v}$ & $\boldsymbol{s}$ & $\boldsymbol{a_1}$ & $\boldsymbol{a_2}$ & $\boldsymbol{a_3}$ & $\boldsymbol{a_4}$ & $\boldsymbol{a_5}$ & $\boldsymbol{a_6}$ & $\boldsymbol{r}=\boldsymbol{u}\wedge\boldsymbol{v}$\\\hline\hline
T&T&T&T&T&T&F&T&F&T\\\hline
T&F&T&T&\multicolumn{2}{c|}{TF or FT}&T&F&T&F\\\hline
F&T&T&T&\multicolumn{2}{c|}{TF or FT}&T&F&T&F\\\hline
F&F&F&F&T&T&T&F&T&F\\\hline
\end{tabular}
\label{tab:andtable}
\end{table}

%%%%%%%%%%%%%%%%%%%%%%%%%%%%%%%%%%%%%%%%%%%%%%%%%%%
%%%%% 4. PP-hardness of Minesweeper revisited %%%%%
%%%%%%%%%%%%%%%%%%%%%%%%%%%%%%%%%%%%%%%%%%%%%%%%%%%

\section{PP-hardness of Minesweeper revisited}

\begin{figure}[ht]
\centering
\begin{tikzpicture} [hexa/.style= {shape=regular polygon,regular polygon sides=6,minimum size=0.7cm, draw,inner sep=0,anchor=south,rotate=30}] 

% fill in some colors

\foreach \j in {2,3,5,6,8,9,11,12} {\node [shape=regular polygon,regular polygon sides=6,minimum size=0.7cm,fill=lightgray!50!white,inner sep=0,anchor=south,rotate=30] at ({(2+\j/2-0.5)*sin(60)*0.7},{2.25*0.7+\j*0.75*0.7}) {};} 
\foreach \i in {3,4,6,7} {\node [shape=regular polygon,regular polygon sides=6,minimum size=0.7cm,fill=lightgray!50!white,inner sep=0,anchor=south,rotate=30] at ({(\i+7/2-0.5)*sin(60)*0.7},{2.25*0.7+7*0.75*0.7}) {};} 
{\node [shape=regular polygon,regular polygon sides=6,minimum size=0.7cm,fill=lightgray!50!white,inner sep=0,anchor=south,rotate=30] at ({(3+6/2-0.5)*sin(60)*0.7},{2.25*0.7+6*0.75*0.7}) {};} 
{\node [shape=regular polygon,regular polygon sides=6,minimum size=0.7cm,fill=lightgray!50!white,inner sep=0,anchor=south,rotate=30] at ({(4+5/2-0.5)*sin(60)*0.7},{2.25*0.7+5*0.75*0.7}) {};} 

% generating nodes
     
\foreach \j in {1,...,14}{
  \node[hexa] (h0;\j) at ({(0+\j/2-0.5)*sin(60)*0.7},{2.25*0.7+\j*0.75*0.7}) {};}
\foreach \j in {0,...,14}{ 
  \node[hexa] (h1;\j) at ({(1+\j/2-0.5)*sin(60)*0.7},{2.25*0.7+\j*0.75*0.7}) {};}
\foreach \j in {0,...,13}{ 
  \foreach \i in {2,3}{  
  \node[hexa] (h\i;\j) at ({(\i+\j/2-0.5)*sin(60)*0.7},{2.25*0.7+\j*0.75*0.7}) {};}  }
\foreach \j in {0,...,12}{ 
  \node[hexa] (h4;\j) at ({(4+\j/2-0.5)*sin(60)*0.7},{2.25*0.7+\j*0.75*0.7}) {};}
\foreach \j in {3,...,9}{ 
  \foreach \i in {5,6}{  
  \node[hexa] (h\i;\j) at ({(\i+\j/2-0.5)*sin(60)*0.7},{2.25*0.7+\j*0.75*0.7}) {};}  }
\foreach \j in {5,...,9}{ 
  \foreach \i in {7,8}{  
  \node[hexa] (h\i;\j) at ({(\i+\j/2-0.5)*sin(60)*0.7},{2.25*0.7+\j*0.75*0.7}) {};}  }
\foreach \j in {5,...,8}{  
  \node[hexa] (h9;\j) at ({(9+\j/2-0.5)*sin(60)*0.7},{2.25*0.7+\j*0.75*0.7}) {};}
\foreach \j in {5,6}{ 
  \node[hexa] (h10;\j) at ({(10+\j/2-0.5)*sin(60)*0.7},{2.25*0.7+\j*0.75*0.7}) {};}
\foreach \j in {-1}{ 
  \foreach \i in {3,4}{ 
  \node[hexa] (h\i;\j) at ({(\i+\j/2-0.5)*sin(60)*0.7},{2.25*0.7+\j*0.75*0.7}) {};}  }

% Putting contents in the nodes
 
\foreach \k in {1,14}  {\node at (h0;\k) {\begin{turn}{60}$\cdots$\end{turn}};}
\foreach \k in {3,4,6,7,9,10,12,13}  {\node at (h0;\k) {1};}
\foreach \k in {2,5,8,11}  {\node at (h0;\k) {2};}

\foreach \k in {0,14}  {\node at (h1;\k) {\begin{turn}{60}$\cdots$\end{turn}};}
\foreach \k in {1,2,4,5,7,8,10,11,13}  {\node [circle,draw,fill=gray,inner sep=3.5] at (h1;\k) {};}
\foreach \k in {3,6,9,12}  {\node at (h1;\k) {3};}

\foreach \k in {0,13}  {\node at (h2;\k) {\begin{turn}{60}$\cdots$\end{turn}};}
\foreach \k in {1,4,10}  {\node at (h2;\k) {5};}
\foreach \k in {7}  {\node at (h2;\k) {4};}
\foreach \k in {2,5,8,11}  {\node at (h2;\k) {$\boldsymbol{s'}$};}
\foreach \k in {3,6,9,12}  {\node at (h2;\k) {$\boldsymbol{s}$};}

\foreach \k in {3,4}  {\node at (h\k;-1) {\begin{turn}{60}$\cdots$\end{turn}};}

\foreach \k in {0,1,3,4,9,10,12}  {\node [circle,draw,fill=gray,inner sep=3.5] at (h3;\k) {};}
\foreach \k in {13}  {\node at (h3;\k) {\begin{turn}{60}$\cdots$\end{turn}};}
\foreach \k in {2,11}  {\node at (h3;\k) {3};}
\foreach \k in {5,8}  {\node at (h3;\k) {4};}
\foreach \k in {7}  {\node at (h3;\k) {$\boldsymbol{x'}$};}
\foreach \k in {6}  {\node at (h3;\k) {$\boldsymbol{x}$};}

\foreach \k in {12}  {\node at (h4;\k) {\begin{turn}{60}$\cdots$\end{turn}};}
\foreach \k in {1,2,10,11}  {\node at (h4;\k) {1};}
\foreach \k in {0}  {\node at (h4;\k) {2};}
\foreach \k in {3}  {\node at (h4;\k) {3};}
\foreach \k in {6,9}  {\node at (h4;\k) {4};}
\foreach \k in {4,8}  {\node [circle,draw,fill=gray,inner sep=3.5] at (h4;\k) {};}
\foreach \k in {5}  {\node at (h4;\k) {$\boldsymbol{x'}$};}
\foreach \k in {7}  {\node at (h4;\k) {$\boldsymbol{x}$};}

\foreach \k in {4,5,6,8}  {\node [circle,draw,fill=gray,inner sep=3.5] at (h5;\k) {};}
\foreach \k in {9}  {\node at (h5;\k) {1};}
\foreach \k in {3}  {\node at (h5;\k) {2};}
\foreach \k in {7}  {\node at (h5;\k) {5};}

\foreach \k in {3,9}  {\node at (h6;\k) {1};}
\foreach \k in {4}  {\node at (h6;\k) {2};}
\foreach \k in {5,8}  {\node at (h6;\k) {3};}
\foreach \k in {7}  {\node at (h6;\k) {$\boldsymbol{x'}$};}
\foreach \k in {6}  {\node [circle,draw,fill=gray,inner sep=3.5] at (h6;\k) {};}

\foreach \k in {5}  {\node at (h7;\k) {1};}
\foreach \k in {9}  {\node at (h7;\k) {2};}
\foreach \k in {6}  {\node at (h7;\k) {3};}
\foreach \k in {7}  {\node at (h7;\k) {$\boldsymbol{x}$};}
\foreach \k in {8}  {\node [circle,draw,fill=gray,inner sep=3.5] at (h7;\k) {};}

\foreach \k in {6,8}  {\node [circle,draw,fill=gray,inner sep=3.5] at (h8;\k) {};}
\foreach \k in {5}  {\node at (h8;\k) {1};}
\foreach \k in {7}  {\node at (h8;\k) {5};}
\foreach \k in {9}  {\node at (h8;\k) {$\cdots$};}

\foreach \k in {7,8}  {\node at (h9;\k) {$\cdots$};}
\foreach \k in {5}  {\node at (h9;\k) {2};}
\foreach \k in {6}  {\node [circle,draw,fill=gray,inner sep=3.5] at (h9;\k) {};}

\foreach \k in {5,6}  {\node at (h10;\k) {$\cdots$};}

% arrows

{\node at ({2.3*0.565},{10.7*0.565}) {$\boldsymbol{s}$};}
{\node at ({3*0.565},{12*0.565}) {\begin{turn}{60}$\xrightarrow{\hspace*{1cm}}$\end{turn}};}

{\node at ({12*0.565},{7*0.565}) {$\boldsymbol{x}\xrightarrow{\hspace*{1cm}}$};}

\end{tikzpicture}
\caption{$x$ can be revealed when $s$ is known}\label{PP}
\end{figure}
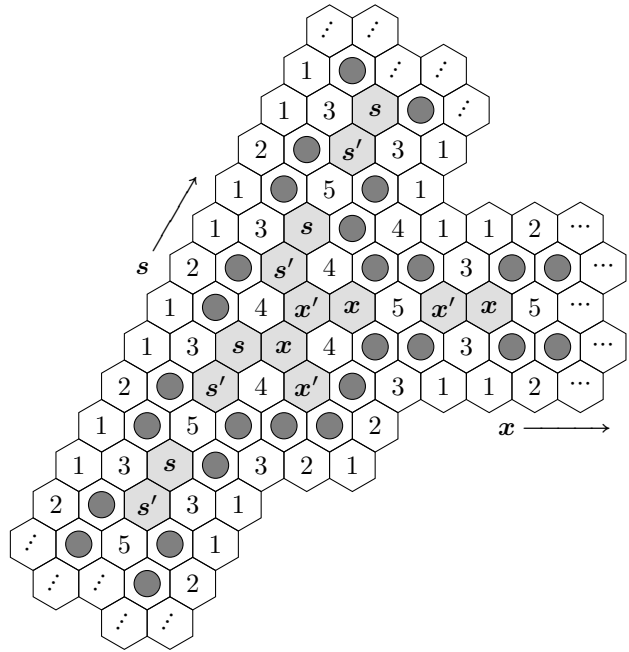

Bondt \cite{bondt12} showed that Minesweeper is PP-hard. First, he proved that weak MAJSAT is PP-complete, and then wired back the output of the circuit to the starting points of the inputs, in such a way that these inputs can be revealed when the output is known. In figure~\ref{PP}, I give such an example with an improved wire.

\vskip 1em

%%%%%%%%%%%%%%%%%%%%%%%%%%%%%%%%%
% Figure 14 : hexagonal PP hard %
%%%%%%%%%%%%%%%%%%%%%%%%%%%%%%%%%

Since $\boldsymbol{x}$ is independent from $\boldsymbol{s}$, Bondt \cite{bondt12} rounded the number $\vartheta$ such that
$$ \reflectbox{\textsf{R}}x_1\reflectbox{\textsf{R}}x_2\cdots\reflectbox{\textsf{R}}x_n\,\mathrm{Pr}\left[ f\left( x_1,x_2,\cdots,x_n \right) \right]=\vartheta $$
to either 0 or 1, with a rounding error of at most 0.5. The symbol \reflectbox{\textsf{R}} means a random quantifier \cite{papa94}. \reflectbox{\textsf{R}}$x_1$ implies a random choice, that is, the probability of true equals to 0.5 for true value for $x_1$. As we have seen above, using an improved wire, we can also say that a hexagonal Minesweeper is PP-hard.

%%%%%%%%%%%%%%%%%%%%%%
%%%%% References %%%%%
%%%%%%%%%%%%%%%%%%%%%%

\end{document}